\documentclass[a4paper,11pt]{article}

\pdfoutput=1

\usepackage{amsmath,amssymb,mathtools}
\usepackage{color}
\usepackage{graphicx}
\usepackage{subfigure}
\usepackage{cite}
\usepackage[colorlinks=true,linkcolor=blue, citecolor=red, urlcolor=blue, bookmarks]{hyperref}
\usepackage{multirow,makecell}
\usepackage{textcomp}
\usepackage{wasysym}

\usepackage{ulem}

\usepackage{simpler-wick} 

\usepackage[text={17.1cm,24.6cm},centering]{geometry} 

\newcommand{\mr}[2]{\multirow{#1}{*}{#2}}

\numberwithin{equation}{section}

\def \be {\begin{equation}}
\def \ee {\end{equation}}
\def \ba {\begin{array}}
\def \ea {\end{array}}
\def \bea{\begin{eqnarray}}
\def \eea{\end{eqnarray}}
\def \nn {\nonumber}

\def \a {\alpha}

\def \c {\gamma}

\def \d {\delta}
\def \D {\Delta}

\def \ve {\varepsilon}

\def \s {\sigma}

\def \r {\rho}

\def \vph {\varphi}

\def \z {\zeta}

\def \cA {\mathcal A}
\def \cB {\mathcal B}
\def \cC {\mathcal C}
\def \cD {\mathcal D}

\def \cF {\mathcal F}

\def \cK {\mathcal K}

\def \cM {\mathcal M}
\def \cN {\mathcal N}
\def \cO {\mathcal O}

\def \cQ {\mathcal Q}
\def \cR {\mathcal R}
\def \cS {\mathcal S}

\def \cX {\mathcal X}

\def \rC {\mathrm C}
\def \rD {\mathrm D}

\def \mO {\mathcal O}

\def \mX {\mathcal X}

\def \p {\partial}

\def \f {\frac}

\def \lt {\left}
\def \rt {\right}

\def \lra {\leftrightarrow}
\def \sr {\sqrt}

\def \pp {\propto}
\def \inf {\infty}

\def \lag {\langle}
\def \rag {\rangle}

\def \ep {\mathrm{e}}
\def \ii {\mathrm{i}}

\def \tr {\textrm{tr}}

\def \and {{~\textrm{and}~}}

\def \SL {{\textrm{SL}}}

\begin{document}


\title{
\textbf{Excited state R\'enyi entropy and subsystem distance in two-dimensional non-compact bosonic theory.\\Part II. Multi-particle states}
}
\author{
Jiaju Zhang$^{1,2}$ 
and
M. A. Rajabpour$^{3}$ 
}
\date{}
\maketitle
\vspace{-12mm}
\begin{center}
{\it
$^{1}$Center for Joint Quantum Studies and Department of Physics, School of Science,\\
Tianjin University, 135 Yaguan Road, Tianjin 300350, China\\\vspace{1mm}
$^{2}$SISSA and INFN, Via Bonomea 265, 34136 Trieste, Italy\\\vspace{1mm}
$^{3}$Instituto de Fisica, Universidade Federal Fluminense,\\
Av. Gal. Milton Tavares de Souza s/n, Gragoat\'a, 24210-346, Niter\'oi, RJ, Brazil
}
\vspace{10mm}
\end{center}

\begin{abstract}
  We study the excited state R\'enyi entropy and subsystem Schatten distance in the two-dimensional free massless non-compact bosonic field theory, which is a conformal field theory. The discretization of the free non-compact bosonic theory gives the harmonic chain with local couplings. We consider the field theory excited states that correspond to the harmonic chain states with excitations of more than one quasiparticle, which we call multi-particle states. This extends the previous work by the same authors to more general excited states. In the field theory we obtain the exact R\'enyi entropy and subsystem Schatten distance for several low-lying states. We obtain short interval expansion of the R\'enyi entropy and subsystem Schatten distance for general excited states, which display different universal scaling behaviors in the gapless and extremely gapped limits of the non-compact bosonic theory. In the locally coupled harmonic chain we calculate numerically the excited state R\'enyi entropy and subsystem Schatten distance using the wave function method. We find excellent matches of the analytical results in the field theory and numerical results in the gapless limit of the harmonic chain. We also make some preliminary investigations of the R\'enyi entropy and the subsystem Schatten distance in the extremely gapped limit of the harmonic chain.
\end{abstract}

\baselineskip 18pt
\thispagestyle{empty}
\newpage

\tableofcontents


\section{Introduction}

Quantum entanglement plays a key role in better understanding of the quantum many-body systems and the quantum field theories (QFTs) \cite{Amico:2007ag,Eisert:2008ur,calabrese2009entanglement,Laflorencie:2015eck,Witten:2018lha}.
For a quantum system in a pure state $|\cX\rag$ with normalization $\a_\cX=\lag \cX | \cX \rag$, the density matrix of the total system is
\be
\r_\cX = \f{1}{\a_\cX} | \cX \rag \lag \cX |.
\ee
One can then divide the whole system into the subsystem $A$ and its complement $B$.
Tracing out the degrees of freedom of $B$, one obtains the reduced density matrix (RDM) $\r_{A,\cX}=\tr_B\r_{A,\cX}$.
Then the entanglement entropy is the von Neumann entropy of the RDM
\be
S_{A,\cX} = - \tr_A ( \r_{A,\cX} \log  \r_{A,\cX} ),
\ee
which is often calculated as the $n \to 1$ limit of the R\'enyi entropy
\be
S_{A,\cX}^{(n)} = - \f{1}{n-1} \log \tr_A\r_{A,\cX}^n.
\ee
When the R\'enyi entropy $S_{A,G}^{(n)}$ of the ground state $|G\rag$ is known, one can use $\cF_{A,\cX}^{(n)}$ to denote the excited state R\'enyi entropy
\be
S_{A,\cX}^{(n)} = S_{A,G}^{(n)} - \f{1}{n-1} \log \cF_{A,\cX}^{(n)}.
\ee
More explicitly, we have
\be
\cF_{A,\cX}^{(n)} = \f{\tr_A \r_{A,\cX}^n}{\tr_A \r_{A,G}^n},
\ee
where $\r_{A,G}$ is the ground state RDM.

The low energy behavior of many one-dimensional critical quantum chains in the continuum limit can be described by two-dimensional (2D) conformal field theories (CFTs) \cite{henkel1999conformal}. It is interesting to compare the entanglement entropy and R\'enyi entropy in CFTs with those in the corresponding critical quantum chains.
The entanglement entropy and R\'enyi entropy have been studied extensively in the ground state and the excited states of various QFTs and quantum chains
\cite{Bombelli:1986rw,Srednicki:1993im,Callan:1994py,Holzhey:1994we,peschel1999density,Peschel:1999DensityMatrices,%
Chung2000Densitymatrix,chung2001density,Vidal:2002rm,peschel2003calculation,Latorre:2003kg,jin2004quantum,Korepin:2004zz,Plenio:2004he,Calabrese:2004eu,%
Cramer:2005mx,Casini:2005rm,Casini:2005zv,Casini:2009sr,Calabrese:2009qy,peschel2009reduced,Alba:2009th,Alcaraz:2011tn,peschel2012special,Berganza:2011mh,%
Pizorn2012Universality,Essler2013ShellFilling,Berkovits2013Twoparticle,Taddia:2013Entanglement,%
Storms2014Entanglement,Palmai:2014jqa,Calabrese:2014Entanglement,Molter2014Bound,Taddia:2016dbm,Castro-Alvaredo:2018dja,Castro-Alvaredo:2018bij,Murciano:2018cfp,%
Castro-Alvaredo:2019irt,Castro-Alvaredo:2019lmj,Jafarizadeh:2019xxc,Capizzi:2020jed,You:2020osa,Haque:2020Entanglement,Wybo:2020fiz}.
Especially, the R\'enyi entropy of one single interval of the ground state of a one-dimensional gapless system takes a universal logarithmic formula  \cite{Holzhey:1994we,Vidal:2002rm,jin2004quantum,Korepin:2004zz,Calabrese:2004eu}.

In quantum information theory, it is often important to know the quantitative difference between two density matrices, especially for two subsystem RDMs.
The subsystem distance has been studied in various circumstances, see for example
\cite{fagotti2013reduced,Basu:2017kzo,He:2017txy,Zhang:2019wqo,Mendes-Santos:2019tmf,Zhang:2019itb,Zhang:2020mjv,Zhang:2019kwu,%
Suzuki:2019xdq,Kusuki:2019hcg,Arias:2020sgz}.
There are many definitions of the distance between two density matrices \cite{nielsen2010quantum,hayashi2017quantum,watrous2018theory}.
In this paper we will focus on the Schatten distance between two RDMs $\r_A$ and $\s_A$ normalized as
\be
D_n( \r_A,\s_A ) = \Big( \f{\tr_A| \r_A - \s_A |^n}{2\tr_A\r_{A,G}^n} \Big)^{1/n}.
\ee
We note that although the above quantity is not a proper measure of distance for generic $n$'s, especially when the subsystem is large, there are at least a couple of good reasons to study it.
First of all in the limit of $n\to1$ it leads to trace distance
\be
D_1( \r_A,\s_A ) = \f12 \tr_A| \r_A - \s_A |,
\ee
which is an important measure of distance.
Secondly it seems that the Schatten distance for even $n$'s capture many properties that we expect for the trace distance \cite{Zhang:2020mjv}.

The present paper is the following work of \cite{Zhang:2020ouz}, where we have calculated the second R\'enyi entropy and the second subsystem Schatten distance in the excited states produced by the current and its derivatives in the 2D free massless non-compact bosonic field theory, which is a 2D CFT with central charge $c=1$, and found perfect matches of the analytical CFT results with the corresponding numerical results in the gapless limit of the harmonic chain with local interactions.
The excited state produced by the current or its derivatives in the massless bosonic theory corresponds to the excited state of one single quasiparticle in the gapless harmonic chain, which we call the single-particle state.
The formulas in \cite{Zhang:2020ouz} only apply to the second R\'enyi entropy and the second Schatten distance among the single-particle states.
In this paper we will consider more general excited states in the 2D massless bosonic theory that correspond to the excited states of more than one quasiparticle in the harmonic chain, which we call the multi-particle states.
In the locally coupled harmonic chain, we will construct the wave function of the general multi-particle states and calculate the R\'enyi entropy and Schatten distance using the full-fledged wave function method \cite{Castro-Alvaredo:2018dja,Castro-Alvaredo:2018bij}. In the massless noncompact bosonic theory, we need to consider not only the current but also quasiparticle operators with larger conformal weights.
Besides the second R\'enyi entropy and the second Schatten distance, we will also calculate the third and the fourth R\'enyi entropies and the fourth Schatten distance.

In the 2D free massless non-compact bosonic theory, we will calculate the exact R\'enyi entropy in several low-lying excited states following \cite{Alcaraz:2011tn,Berganza:2011mh,Essler2013ShellFilling,Calabrese:2014Entanglement,Taddia:2016dbm}.
For general excited states, we will calculate the leading order of the short interval expansion of the R\'enyi entropy using twist operators \cite{Calabrese:2004eu,Cardy:2007mb,Calabrese:2009qy} and their operator product expansion (OPE) \cite{Headrick:2010zt,Calabrese:2010he,Rajabpour:2011pt,Chen:2013kpa}.
Similarly, we will calculate the exact value \cite{Zhang:2019wqo,Zhang:2019itb} and short interval expansion \cite{Basu:2017kzo,He:2017txy,Arias:2020sgz} of the Schatten distance in the 2D massless bosonic theory.
In the locally coupled harmonic chain, we will use the wave function method \cite{Castro-Alvaredo:2018dja,Castro-Alvaredo:2018bij} to calculate numerically the R\'enyi entropy and subsystem Schatten distance.
The analytical R\'enyi entropy and Schatten distance in the massless bosonic theory match perfectly with the numerical results in the gapless limit of the harmonic chain.

Compared to our previous work \cite{Zhang:2020ouz}, in this paper we develop calculation methods in both the field theory and the harmonic chain.
In field theory, besides the states generated by the current operator $J$ and its derivatives we also consider the states generated by the stress tensor $T$ and its first derivative, as well as the quasiprimary operator $\cK$, defined in (\ref{cKdef}), with conformal weight 3. In the harmonic chain, we generalize the wave function method in \cite{Castro-Alvaredo:2018dja,Castro-Alvaredo:2018bij} by giving the trace of the multiplication of arbitrary number of reduced density matrices of the most general energy eigenstates. In \cite{Zhang:2020ouz} we have only calculated the second R\'enyi entropy and the second Schatten distance, for which we only need to consider four-point functions in field theory. In this paper we also calculate the third and the fourth R\'enyi entropies and the fourth Schatten distance, and we need to consider up to eight-point functions.

In this paper we also investigate the short interval expansion of the R\'enyi entropy and the Schatten distance in the gapless and extremely gapped limits of the harmonic chain, and find different universal scaling behaviors for the leading order term. In the gapless limit, the leading order R\'enyi entropy and Schatten distance depend on energies of the states, while in the extremely gapped limit, the leading order R\'enyi entropy and Schatten distance depend on the numbers of excited quasiparticles of the states. The generalization from single-particle states in \cite{Zhang:2020ouz} to multi-particle states in this paper is essential to obtain these universal short interval scaling beharviors.

The remaining part of the paper is organized as follows:
In section~\ref{secCFT}, we present the useful details of the 2D free massless non-compact bosonic field theory, and we also calculate the products of RDMs in the excited states of some holomorphic operators, which are the main ingredients to calculate the R\'enyi entropy and Schatten distance.
In section~\ref{secHC}, we review the excited state wave functions in the locally coupled harmonic chain and the wave function method to calculate the products of RDMs.
In section~\ref{secRE} we compare the analytical field theory results and the numerical harmonic chain results of the R\'enyi entropy.
In section~\ref{secSD} we demonstrate the analytical and numerical results of the Schatten distance.
We conclude with discussions in section~\ref{secDis}.
In appendix~\ref{appKKKK} we give two alternative ways to derive a useful four-point correlation function.
In appendix~\ref{appPro} we collect the field theory results of the products of the RDMs.

\section{Free massless non-compact bosonic theory}\label{secCFT}

In this section we first present some useful details of the quasiprimary operators and their multi-point correlation functions in the 2D free massless non-compact bosonic field theory.
One can find review of the basics of the 2D free massless bosonic theory in \cite{DiFrancesco:1997nk,Blumenhagen:2009zz}.
For small conformal weights, the results are well-known, while at higher conformal weights we need to derive them following the basic rules.
The excited states in the CFT are constructed by applying the quasiprimary operators and their derivatives on the ground state.
Then we calculate the products of the excited state RDMs that are the basic ingredients to calculate the R\'enyi entropy and the Schatten distance.

\subsection{Quasiprimary operators and correlation functions}

In this subsection we will focus on the holomorphic sector of the theory on a complex plane with coordinates $z$, $\bar z$. The anti-holomorphic sector is similar to the holomorphic sector.

The scalar field can be decomposed as the sum of the holomorphic and the anti-holomorphic parts $\phi(z,\bar z) = \vph(z) + \bar\vph(\bar z)$.
All the holomorphic quasiprimary operators that we consider in this paper can be constructed using the current $J(z)=\ii\p\vph(z)$, which is a primary operator with conformal weight $h_J=1$.
In the identity conformal family, the stress tensor is
\be
T(z) = \f12 (JJ)(z),
\ee
where $(\cdots)$ means the normal ordering.
The stress tensor $T$ is a quasiprimary operator with conformal weight $h_T=2$.
In the conformal family of $J$, there is also the quasiprimary operator at level 2
\be \label{cKdef}
\cK = (TJ) - \f12 \p^2 J = \f12 (J(JJ)),
\ee
with conformal weight $h_\cK=3$.

One can apply an operator $\cX$, which is a quasiprimary operator or the derivatives of a quasiprimary operator, on the ground state and get the excited state on a cylinder
\be
|\cX\rag = \cX(0) |G\rag,
\ee
where $|G\rag$ is the ground state.
We write the normalization factor $\a_\cX = \lag \cX | \cX \rag$.
For the introduced states the normalization factors are
\be
\a_T = \f12, ~~ \a_J=1, ~~ \a_\cK=\f32,
\ee
and we also need
\be
\a_{\p J}=2, ~~ \a_{\p^2 J}=12, ~~ \a_{\p T} = 2.
\ee
For two orthogonal holomorphic quasiprimary operators $\cX_1$ and $\cX_2$, the two-point correlation function on a plane is%
\footnote{Note that we have not normalized the quasi-primary operators and the normalization convention would affect the values of the structure constants.
We will also need to normalize the states and the density matrices properly.}
\be
\lag \cX_1(z_1) \cX_2(z_2) \rag_\rC = \f{\a_{\cX_1} \d_{\cX_1\cX_2} }{z_{12}^{2h_{\mX_1}}}.
\ee
We have defined $z_{12}=z_1-z_2$ and we will also use similar notation below.
For three quasiprimary operators $\cX_1$, $\cX_2$ and $\cX_3$, the three-point correlation function is
\be
\lag \cX_1(z_1) \cX_2(z_2) \cX_3(z_3) \rag_\rC = \f{C_{\cX_1\cX_2\cX_3}}
                                               {z_{12}^{h_{\cX_1}+h_{\cX_2}-h_{\cX_3}}
                                                z_{13}^{h_{\cX_1}+h_{\cX_3}-h_{\cX_2}}
                                                z_{23}^{h_{\cX_2}+h_{\cX_3}-h_{\cX_1}}},
\ee
with the structure constant $C_{\cX_1\cX_2\cX_3}$.
In particular, we will need
\bea
&& C_{TTT}=1, ~~ C_{TJJ}=1, ~~ C_{TJ\cK}=\f32, ~~ C_{T\cK\cK}=\f92, \nn\\
&& C_{TTJ}=C_{TT\cK}=C_{JJJ}=C_{JJ\cK}=C_{J\cK\cK}=C_{\cK\cK\cK}=0.
\eea

We write the correlation function of four holomorphic quasiprimary operators $\cX_1$, $\cX_2$, $\cX_3$ and $\cX_4$ as
\be
\lag \cX_1(z_1) \cX_2(z_2) \cX_3(z_3) \cX_4(z_4)  \rag_\rC = \f{1}{z_{12}^{h_{\cX_1}+h_{\cX_2}} z_{34}^{h_{\cX_3}+h_{\cX_4}} }
                                                      \Big(\f{z_{24}}{z_{14}}\Big)^{h_{\cX_1}-h_{\cX_2}}
                                                      \Big(\f{z_{14}}{z_{13}}\Big)^{h_{\cX_3}-h_{\cX_4}}
                                                      F_{\cX_1\cX_2\cX_3\cX_4}(r),
\ee
where the function $F_{\cX_1\cX_2\cX_3\cX_4}(r)$ is solely dependent on the cross ratio $r=\f{z_{12}z_{34}}{z_{13}z_{24}}$.
The multi-point correlation functions can be calculated from the lower-point correlation functions using
\be \label{lagcX1cX2cdotscXmrag}
\lag \cX_1(z_1) \cX_2(z_2) \cdots \cX_m(z_m) \rag_\rC = \sum_{i=2}^m \lag \wick{ \c \cX_1(z_1) \cX_2(z_2) \cdots \c \cX_i(z_i) \cdots \cX_m(z_m) } \rag_\rC,
\ee
with the Wick contraction denoting the singular terms in the OPE.
Note that we only consider bosonic operators in this paper and so we could switch freely the positions of the operators in the correlation functions.
The useful OPEs are
\bea \label{OPEs}
&& \wick{\c T(z) \c T(w)} = \f{1}{2(z-w)^4} + \f{2T(w)}{(z-w)^2} + \f{\p T(w)}{z-w}, \nn\\
&& \wick{\c T(z) \c J(w)} = \f{J(w)}{(z-w)^2} + \f{\p J(w)}{z-w}, \nn\\
&& \wick{\c T(z) \c \cK(w)} = \f{3J(w)}{2(z-w)^4} + \f{3\cK(w)}{(z-w)^2} + \f{\p\cK(w)}{z-w}, \nn\\
&& \wick{\c J(z) \c J(w)} = \f{1}{(z-w)^2}, \nn\\
&& \wick{\c J(z) \c \cK(w)} = \f{3T(w)}{(z-w)^2}, \nn\\
&& \wick{\c \cK(z) \c \cK(w)} = \f{3}{2(z-w)^6}
                              + \f{9}{(z-w)^4} \Big[ T(w) + \f12 (z-w) \p T(w) \nn\\
&& \phantom{\wick{\c \cK(z) \c \cK(w)} =}
                              + \f{3}{20} (z-w)^2 \p^2 T(w)
                              + \f{1}{30} (z-w)^3 \p^3 T(w) \Big] \nn\\
&& \phantom{\wick{\c \cK(z) \c \cK(w)} =}
                              + \f{7}{(z-w)^2} \Big[ \cA(w) + \f12 (z-w) \p \cA(w) \Big] \nn\\
&& \phantom{\wick{\c \cK(z) \c \cK(w)} =}
                              + \f{1}{2(z-w)^2} \Big[ \cB(w) + \f12 (z-w) \p \cB(w) \Big].
\eea
In the above $\cK\cK$ OPE, there are orthogonal quasiprimary operators
\bea \label{cAcBdef}
&& \cA = (TT) - \f{3}{10} \p^2 T
       = \f{1}{4} (J(J(JJ))) - \f{3}{10} (\p J \p J) + \f{1}{5} ( \p^2JJ ), \nn\\
&& \cB = (J(J(JJ))) + \f{3}{2} (\p J \p J) - ( \p^2JJ ),
\eea
with the following conformal weights, normalization factors and structure constants
\bea
&& h_\cA=4, ~~ \a_\cA = \f{27}{10}, ~~ C_{\cK\cK\cA} = \f{189}{10}, \nn\\
&& h_\cB=4, ~~ \a_\cB = 54, ~~ C_{\cK\cK\cB} = 27.
\eea
The operator $\cA$ is a descendant operator in the identity conformal family, and the operator $\cB$ is not only quasiprimary but also primary.
In the conformal family of the current $J$, there is a quasiprimary operator at level 3, i.e. conformal weight 4,
\be
\cN = (\p T J) - 2 (T \p J) + \f12 \p^3 J =0,
\ee
which is a null operator and does not belong to the spectrum.
Then using the formula (\ref{lagcX1cX2cdotscXmrag}) and the OPEs (\ref{OPEs}) one can get the four-point correlation functions on a plane
\bea \label{FXYZWr}
&& F_{TTTT}(r) = \f{(1 - r + r^2) (1 - 3 r + 10 r^2 - 15 r^3 + 10 r^4 - 3 r^5 + r^6)}{4 (1 - r)^4}, \nn\\
&& F_{TTJJ}(r) = \f{1 - 2 r + 5 r^2 - 4 r^3 + 2 r^4}{2 (1 - r)^2},\nn\\ ~~
&&   F_{TTJ\cK}(r) = \f{3 r^2 (2 - 4 r + 4 r^2 - 2 r^3 + r^4)}{2 (1 - r)^2}, \nn\\
&& F_{TT\cK\cK}(r) = \f{3 (1 - 4 r + 18 r^2 - 40 r^3 + 61 r^4 - 60 r^5 + 36 r^6 - 12 r^7 +3 r^8)}{4 (1 - r)^4}, \nn\\
&& F_{JJJJ}(r) = \f{(1 - r + r^2)^2}{(1 - r)^2}, \nn\\~~
&&   F_{JJJ\cK}(r) = 3 r^2, \nn\\
&& F_{JJ\cK\cK}(r) = \f{3 (1 - 2 r + 7 r^2 - 6 r^3 + 3 r^4)}{2 (1 - r)^2}, \nn\\~~
&&   F_{J\cK\cK\cK}(r) = \f{27 r^2 (1 - r + r^2)^2}{2 (1 - r)^4}, \nn\\
&& F_{\cK\cK\cK\cK}(r) = \f{9}{4(1-r)^6} ( 1 - 6 r + 33 r^2 - 110 r^3 + 258 r^4 - 438 r^5 + 525 r^6 \nn\\
&& \phantom{F_{\cK\cK\cK\cK}(r) =}
   - 438 r^7 + 258 r^8 - 110 r^9 + 33 r^{10} - 6 r^{11} + r^{12}).
\eea
One could also obtain the four-point correlation function $\lag\cK\cK\cK\cK\rag_\rC$ using a couple of other methods that we discuss in appendix~\ref{appKKKK}.

The multi-point correlation functions of solely the current $J$ can be calculated easily using the Wick contraction.
In the following subsection, we will need the six-point and eight-point correlation functions
\bea \label{J6J8}
&& \lag J(z_1) J(z_2) \cdots J(z_6) \rag_\rC = \Big( \f{1}{x_{12}^2 x_{34}^2 x_{56}^2} + \rm{perm} \Big)_{15}, \nn\\
&& \lag J(z_1) J(z_2) \cdots J(z_8) \rag_\rC = \Big( \f{1}{x_{12}^2 x_{34}^2 x_{56}^2 x_{78}^2} + \rm{perm} \Big)_{105},
\eea
where ``perm'' denotes the permutation terms. In the six-point and eight-point correlation functions there are totally 15 and 105 terms, respectively.

In the next subsection, we will also need the transformation rules under the general conformal transformation $z \to f(z)$
\bea
&& T(z) = f'^2 T(f) + \f{1}{24}s, \nn\\
&& J(z) = f' J(f), \nn\\
&& \cK(z) = f'^3 \cK(f) + \f{1}{4} s f' J(f),
\eea
with the Schwarzian derivative
\be \label{Schwarzian}
s(z) = \f{f'''(z)}{f'(z)} - \Big( \f{f''(z)}{f'(z)} \Big)^2.
\ee
Note that $J$ is a primary operator, while $T$ and $\cK$ are only quasiprimary operators.
For a transformation $f(z) \in \SL(2,\rC)$, all of the operators, i.e.  $\cX\in\{T,J,\cK\}$, transform as
\be
\cX(z) = f'^{h_\cX} \cX(f).
\ee

The operators in the anti-holomorphic sector can be studied similarly to the holomorphic operators.
In addition, one can also consider the operators that are products of the purely holomorphic and purely anti-holomorphic operators.
The correlation functions and the transformation rules for the purely anti-holomorphic operators and the product operators can be obtained easily from those of the holomorphic operators.
We will not show details here.
We list the holomorphic operators, the anti-holomorphic operators, and some examples of the product operators that we will consider in table~\ref{operators}.
With the six purely holomorphic operators and six purely anti-holomorphic operators, one can construct 36 product operators that we only show some examples in the table.

\begin{table}[t]
  \centering
  \begin{tabular}{|c|c|c|c|c|c|c|}\cline{1-3}\cline{5-7}
  $(h,\bar h)$  & $\cX$            & $K$        && $(h,\bar h)$  & $\cX$                & $K$           \\ \cline{1-3}\cline{5-7}
  (1,0)         & $J$              & 1          && (1,1)         & $J\bar J$            & $1(-1)$       \\ \cline{1-3}\cline{5-7}

  \mr{2}{(2,0)} & $\p J$           & 2          && \mr{2}{(1,2)} & $J\bar\p\bar J$      & $1(-2)$       \\ \cline{2-3}\cline{6-7}
                & $T$              & $1^2$      &&               & $J\bar T$            & $1(-1)^2$     \\ \cline{1-3}\cline{5-7}

                & $\p^2J$          & 3          &&               & $J\bar\p^2\bar J$    & $1(-3)$       \\ \cline{2-3}\cline{6-7}
  (3,0)         & $\p T$           & $1\!\cdot\!2$  && (1,3)         & $J\bar\p\bar T$      & $1(-1)(-2)$   \\ \cline{2-3}\cline{6-7}
                & $\cK$            & $1^3$      &&               & $J\bar\cK$           & $1(-1)^3$     \\ \cline{1-3}\cline{5-7}

  (0,1)         & $\bar J$         & $-1$       && \mr{2}{(2,2)} & $\p J\bar T$         & $2(-1)^2$     \\ \cline{1-3}\cline{6-7}

  \mr{2}{(0,2)} & $\bar\p\bar J$   & $-2$       &&               & $T\bar T$            & $1^2(-1)^2$   \\ \cline{2-3}\cline{5-7}
                & $\bar T$         & $(-1)^2$   && \mr{2}{(2,3)} & $\p J\bar\p^2\bar J$ & $2(-3)$       \\ \cline{1-3}\cline{6-7}

                & $\bar\p^2\bar J$ & $-3$       &&               & $T\bar\cK$           & $1^2(-1)^3$   \\ \cline{2-3}\cline{5-7}
  (0,3)         & $\bar\p\bar T$   & $(-1)(-2)$ && \mr{2}{(3,3)} & $\p T\bar\p^2\bar J$ & $1\!\cdot\!2(-3)$ \\ \cline{2-3}\cline{6-7}
                & $\bar\cK$        & $(-1)^3$   &&               & $\cK\bar\cK$         & $1^3(-1)^3$   \\ \cline{1-3}\cline{5-7}
  \end{tabular}
  \caption{The correspondence of the CFT operators $\cX$ with conformal weight $(h,\bar h)$ and harmonic chain states $K$. We show the holomorphic operators, the anti-holomorphic operators, and the examples of the operators that are products of the holomorphic and anti-holomorphic operators.}
  \label{operators}
\end{table}

\subsection{Trace of products of excited state RDMs}\label{subsecPro}

We consider a length $\ell$ interval $A$ on a cylinder with circumference $L$.
For a holomorphic operator $\cX\in\{G,J,\p J,T,\p^2 J,\p T,\cK\}$, where we use $G$ to denote the identity operator, one can construct the excited state $|\cX\rag=\cX(0)|G\rag$, the density matrix of the whole system and the subsystem RDM
\be
\r_\cX = \f{1}{\a_\cX} |\cX\rag\lag\cX|, ~~
\r_{A,\cX}=\tr_B\r_\cX.
\ee
For two holomorphic operators $\cX_1$ and $\cX_2$ that can be the same or different, we define the product of the RDMs \cite{Alcaraz:2011tn,Berganza:2011mh,Taddia:2016dbm}
\be \label{FAcX1cX2CFT}
\cF_A^{\cX_1,\cX_2} = \f{\tr_A(\r_{A,\cX_1}\r_{A,\cX_2})}{\tr_A\r_{A,G}^2}
                    = \f{1}{\a_{\cX_1}\a_{\cX_2}} \lag \cX_1(\inf_1)\cX_1(0_1)\cX_2(\inf_2)\cX_2(0_2) \rag_{\rC^2},
\ee
which is a four-point correlation function on a two-fold plane $\rC^2$.
The subscripts 1 and 2 in the positions $\inf_1$, $0_1$, $\inf_2$ and $0_2$ are the replica indices.
The two copies of the plane are connected through the interval $[\ep^{-\f{\ii\pi\ell}{L}},\ep^{\f{\ii\pi\ell}{L}}]$, and the two-fold plane with coordinate $z$ can be mapped to a plane with coordinate $\z$ by
\be \label{zz}
\z(z) = \Big( \f{\ep^{-\pi\ii\ell/L} z-1}{z-\ep^{-\pi\ii\ell/L}} \Big)^{1/n},
\ee
with $n=2$.
Then (\ref{FAcX1cX2CFT}) is mapped to a four-point function on an ordinary plane.
Explicitly, we use the results (\ref{FXYZWr}) in the previous subsection and obtain the results of the RDM products that are collected in appendix~\ref{appPro}.

More generally, one can calculate the product of $n$ RDMs
\be \label{cFAcX1cdotscXn}
\cF_A^{\cX_1 \cdots \cX_n} = \f{\tr_A( \r_{A,\cX_1} \cdots \r_{A,\cX_n} )}{\tr_A\r_{A,G}^n}
                           = \f{1}{\a_{\cX_1}\cdots\a_{\cX_n}} \lag \cX_1(\inf_1)\cX_1(0_1) \cdots \cX_n(\inf_n)\cX_n(0_n) \rag_{\rC^n},
\ee
which is a $2n$-point correlation function on an $n$-fold plane $\rC^n$ and can be further mapped to a $2n$-point correlation function on an ordinary plane by (\ref{zz}).
We use (\ref{J6J8}) and calculate some results for $n=3$ and $n=4$ that are collected in appendix~\ref{appPro}.

Some of the results collected in appendix~\ref{appPro} have been derived in \cite{Alcaraz:2011tn,Berganza:2011mh,Essler2013ShellFilling,Calabrese:2014Entanglement,Taddia:2016dbm,Zhang:2020ouz}.
The R\'enyi entropies in the state of current operator $\cF_A^{J,J}$, $\cF_A^{J,J,J}$ and $\cF_A^{J,J,J,J}$ have been already derived in \cite{Alcaraz:2011tn,Berganza:2011mh,Essler2013ShellFilling,Calabrese:2014Entanglement}.
Especially, there is a closed form of the R\'enyi entropy $\cF_{A,J}^{(n)}$ with the general index $n$ derived in \cite{Essler2013ShellFilling,Calabrese:2014Entanglement}.
One can also find the second R\'enyi entropies in the excited states of the stress tensor and its first derivative $\cF_A^{T,T}$ and $\cF_A^{\p T,\p T}$ in \cite{Taddia:2016dbm}.
We have also derived the products of two RDMs among the ground state and states of the current and its derivatives, i.e. $\cF_A^{G,\p^r J}$ and $\cF_A^{\p^r J,\p^s J}$, already in \cite{Zhang:2020ouz}.

\section{Harmonic chain}\label{secHC}

In this section we review the excited states and their wave functions in the locally coupled harmonic chain.
We also review the wave function method \cite{Castro-Alvaredo:2018dja,Castro-Alvaredo:2018bij} to calculate the products of the excited state RDMs.

\subsection{Excited state wave functions}

We consider the discrete version of the 2D free massless non-compact bosonic field theory, which is just the harmonic chain with local interactions
\be
H = \f{1}{2} \sum_{j=1}^L \big[ p_j^2 + m^2 q_j^2 + (q_j-q_{j+1})^2 \big].
\ee
We impose the periodic boundary condition $q_{L+1}=q_L$ and only consider the case that the number of the sites $L$ is an even integer, and so there are the integer momenta
\be
k=1-\f{L}{2},\cdots,-1,0,1,\cdots,\f{L}{2}-1,\f{L}{2}.
\ee
Using the Fourier transformation
\be
q_j = \f{1}{\sr{L}} \sum_k \ep^{-\f{2\pi\ii j k}{L}}\vph_k, ~~
p_j = \f{1}{\sr{L}} \sum_k \ep^{-\f{2\pi\ii j k}{L}}\pi_k,
\ee
the energies of the modes
\be
\ve_k = \sr{m^2+4\sin^2\f{\pi k}{L}},
\ee
and the definitions of the bosonic ladder operators
\be \label{bkbkdag}
b_k=\sr{\f{\ve_k}{2}}\Big( \vph_k + \f{\ii}{\ve_k} \pi_k \Big), ~~
b_k^\dag=\sr{\f{\ve_k}{2}}\Big( \vph_k^\dag - \f{\ii}{\ve_k} \pi_k^\dag \Big),
\ee
the Hamiltonian can be diagonalized as
\be
H = \sum_k \ve_k \Big( b_k^\dag b_k +\f12 \Big).
\ee
The ground state is annihilated by all the lowering operators
\be
b_k |G\rag = 0, ~ k = 1-\f{L}{2},\cdots,\f{L}{2}.
\ee
In the coordinate basis, the ladder operators (\ref{bkbkdag}) are just
\be
b_k=\sr{\f{\ve_k}{2}}\Big( \vph_k + \f{1}{\ve_k} \p_{\vph_k} \Big), ~~
b_k^\dag=\sr{\f{\ve_k}{2}}\Big( \vph_{-k} - \f{1}{\ve_k} \p_{\vph_{-k}} \Big).
\ee
Note that $\vph_{k}^\dag=\vph_{-k}$.
Then the ground state wave function is just
\be
\lag Q | G \rag = \Big( \det\f{W}{\pi} \Big)^{1/4} \ep^{-\f12 Q^T W Q },
\ee
where the coordinates are $Q=(q_1,\cdots,q_L)$ and the $L\times L$ real symmetric matrix $W$ has the entries
\be \label{Wj1Wj2}
W_{j_1j_2} = \f1{L}\sum_k\ve_k\cos\f{2\pi k(j_1-j_2)}{L}.
\ee

The excited states are obtained by applying the raising modes on the ground state.
We may denote the set of excited modes as
\be
K=\{r_{1-\f{L}{2}}, \cdots, r_{\f{L}{2}}  \},
\ee
and the corresponding excited state as
\be \label{stateR}
| K \rag = \Big[ \prod_k \f{(b_k^\dag)^{r_k}}{\sr{r_k!}} \Big] |G\rag.
\ee
We may also equivalently write $K$ as
\be \label{Kk1r1cdotsksrs}
K = k_1^{r_{k_1}} \cdots k_s^{r_{k_s}},
\ee
for the corresponding excited state
\be \label{stateKX}
|K\rag = \f{(b_{k_1}^\dag)^{r_{k_1}} \cdots (b_{k_s}^\dag)^{r_{k_s}}}{\sr{r_{k_1}!\cdots r_{k_s}!}} |G\rag.
\ee
The $L$ different modes excited by $b_k^\dag$ with $k=1-\f{L}{2},\cdots,\f{L}{2}$ can be classified into two types.
The first type are real modes $k=0,\f{L}{2}$, which are their own anti-particles.
The other ones are the complex modes $k=1-\f{L}{2}, \cdots,-1,1,\cdots,\f{L}{2}-1$, where each mode $k$ has the anti-particle mode $-k$.
For $k=0$ or $k=\f{L}{2}$, the wave function is
\be
f_k^r(Q) = \f{\lag Q | k^r \rag}{\lag Q | G \rag} = \f{1}{\sr{2^rr!}} H_r \Big( \f{u_k}{\sr2} \Big),
\ee
and for $k\neq0$ and $k\neq\f{L}{2}$ we have
\be
f_k^{r,s}(Q) = \f{\lag Q | k^r (-k)^s \rag}{\lag Q | G \rag} = \f{1}{\sr{2^{r+s}r!s!}} H_{r,s} \Big( \f{u_k}{\sr2}, \f{u_{-k}}{\sr2} \Big),
\ee
with the definition
\be
u_k = \sr{\f{2\ve_k}{L}} \sum_{j=1}^L \ep^{-\f{2\pi\ii j k}{L}} q_j.
\ee
Note that $(u_k)^*=u_{-k}$.
We have also used the following definitions of the Hermite polynomials
\be
H_r(x) = \ep^{\f{x^2}{2}} ( x - \p_x )^r \ep^{-\f{x^2}{2}}
       = (-)^r\ep^{x^2}\p_x^r \ep^{-x^2},
\ee
and the complex Hermite polynomials
\be \label{polynomials}
H_{r,s}(z,\bar z) = \ep^{z\bar z} ( z - \p_{\bar z} )^r ( \bar z - \p_{z} )^s \ep^{-z\bar z}
                  = (-)^{r+s} \ep^{2z\bar z} \p_{\bar z}^r\p_z^s \ep^{-2z\bar z}.
\ee
Note that all the coefficients of the polynomials $H_r(x)$ and $H_{r,s}(z,\bar z)$ are real.
It is easy to see $H_{r,s}(z,\bar z) = H_{s,r}(\bar z, z)$ and $H_{r,0}(z,\bar z)=(2z)^r$.
A few examples of the Hermite polynomials are:
\be
H_1(x) = 2 x, ~~
H_2(x) = 2 (2 x^2-1), ~~
H_3(x) = 4 x (2 x^2-3), ~~
H_4(x) = 4 (4 x^4-12 x^2+3),
\ee
and the examples of the complex Hermite polynomials are:
\bea
&& H_{1,1}(z,\bar z) = 2 (2 z \bar{z}-1), ~~
   H_{2,1}(z,\bar z) = 8 z (z \bar{z}-1), ~~
   H_{3,1}(z,\bar z) = 8 z^2 (2 z \bar{z}-3), \\
&& H_{2,2}(z,\bar z) = 8 (2 z^2 \bar{z}^2-4 z \bar{z}+1), ~~
   H_{4,1}(z,\bar z) = 32 z^3 (z \bar{z}-2), ~~
   H_{3,2}(z,\bar z) = 16 z (2 z^2 \bar{z}^2-6 z \bar{z}+3). \nn
\eea
For the general excited state (\ref{stateR}), the wave function is
\be \label{fKQ}
f_K(Q) = \f{\lag Q | K \rag}{\lag Q | G \rag} = f_0^{r_0}(Q) f_{L/2}^{r_{L/2}}(Q) \prod_{k=1}^{L/2-1} f_k^{r_k,r_{-k}} ( Q ).
\ee
We see that excited state wave function is generally nontrivial, especially when more than one real modes are excited and/or when both a complex mode and its anti-particle mode are excited.
In the following, we will not consider the states with excitations of the real modes $k=0,\f{L}{2}$.

The correspondence of the modes in the 2D free massless non-compact bosonic field theory and the gapless locally coupled harmonic chain was established in \cite{Zhang:2020ouz} for $k>0$
\be \label{correspondence}
\f{J_{-k}}{\sr{k}} \lra b_k^\dag, ~~
\f{\bar J_{-k}}{\sr{k}} \lra b_{-k}^\dag.
\ee
The excited states produced by the various quasiprimary operators and their derivatives in the field theory considered in this paper can be written in terms of the modes $J_{-k}$ and $\bar J_{-k}$, while the corresponding harmonic chain excited states are written in terms of $b_k^\dag$ and $b_{-k}^\dag$.
The correspondence (\ref{correspondence}) helps the identification of the CFT operators $\cX$ with the harmonic chain states $K$ written in the form of (\ref{Kk1r1cdotsksrs}).
For the holomorphic operators, we get
\bea
|J\rag = J_{-1} |G\rag             &\lra& K = 1, \nn\\
|\p J\rag = J_{-2} |G\rag          &\lra& K = 2, \nn\\
|T\rag = \f12 (J_{-1})^2 |G\rag    &\lra& K = 1^2, \nn\\
|\p^2 J\rag = 2 J_{-3} |G\rag      &\lra& K = 3, \nn\\
|\p T\rag = J_{-1} J_{-2} |G\rag   &\lra& K = 1\!\cdot\!2, \nn\\
|\cK \rag = \f12 (J_{-1})^3 |G\rag &\lra& K = 1^3.
\eea
It is similar for the anti-holomorphic operators and the product operators.
The product of a holomorphic operator and an anti-holomorphic operator is trivial in the CFT, but the corresponding state wave function in the harmonic chain is not necessarily a product and may be nontrivial.
We show the correspondence of the CFT operators $\cX$, including the holomorphic operators, the anti-holomorphic operators, and the examples of the product operators, and harmonic chain states $K$ in the table~\ref{operators}.

\subsection{Wave function method}\label{WFM}

We consider the single interval $A=[1,\ell]$ and its complement $B=[\ell+1,L]$.
Accordingly, we decompose the coordinate $Q=(\cR,\cS)$ with $\cR=(q_1,\cdots,q_\ell)$ and $\cS=(q_{\ell+1},\cdots,q_L)$.
Correspondingly, the $L\times L$ matrix $W$ (\ref{Wj1Wj2}) can be decomposed into $2\times2$ blocks
\be
W = \Big( \ba{cc} \cA & \cB \\ \cC & \cD \ea \Big),
\ee
with the matrices $\cA$, $\cB$, $\cC$ and $\cD$ of respectively $\ell\times\ell$, $\ell\times(L-\ell)$, $(L-\ell)\times\ell$ and $(L-\ell)\times(L-\ell)$ entries.

By replica trick, one defines the coordinate $q_{a,j}$ with the replica index $a=1,2,\cdots,n$.
Then one defines the vector with $nL$ components $\cQ=( \cR_1,\cS_1,\cdots,\cR_n,\cS_n )$ with $\cR_a=(q_{a,1},\cdots,q_{a,\ell})$ and $\cS_a=(q_{a,\ell+1},\cdots,q_{a,L})$.
When the total system is in a general state $\r$, one can write the moments of the RDM $\r_A$ as \cite{Callan:1994py,Castro-Alvaredo:2018bij}
\be
\tr_A \r_{A}^n = \int \rD \cQ
                   \lag \cR_1, \cS_1 | \r | \cR_2, \cS_1 \rag
                   \lag \cR_2, \cS_2 | \r | \cR_3, \cS_2 \rag
                   \cdots
                   \lag \cR_n, \cS_n | \r | \cR_1, \cS_n \rag.
\ee
For the ground state, one gets
\be
\tr_A \r_{A,G}^n = \Big( \det \f{W}{\pi} \Big)^{\f{n}{2}} \int \rD \cQ \ep^{-\f12 \cQ^T \cM_n \cQ}
                 =  \sr{\f{(\det W)^n}{\det\f{\cM_n}{2}} },
\ee
with the $nL\times nL$ matrix $\cM_n$ in the form of $2n\times 2n$ blocks
\be \label{cM}
\cM_n = \lt( \ba{cc|cc|cc|cc}
2\cA & \cB  &        & & & & & \cB \\
\cC  & 2\cD & \cC    & & & & & \\ \hline
     & \cB  & 2\cA   & \cB    & & & & \\
     &      & \cC    & 2\cD   & \cC    & & & \\ \hline
     &      &        & \cB    & \ddots & \ddots &      & \\
     &      &        &        & \ddots & \ddots & \cC  & \\ \hline
     &      &        &        &        & \cB    & 2\cA & \cB \\
\cC  &      &        &        &        &        & \cC  & 2\cD
\ea \rt).
\ee
Note that $\cM_1=2W$, and one check easily the correct normalization $\tr_A\r_{A,G}=1$.
For a general excited state $\r_K$, one gets
\be \label{trArAKn}
\f{\tr_A \r_{A,K}^n}{\tr_A \r_{A,G}^n } 
                                        = \Big\lag\!\!\Big\lag \prod_{a=1}^n [ f_K(\cR_a,\cS_a) f_K^*(\cR_{a+1},\cS_a) ] \Big\rag\!\!\Big\rag,
\ee
with the definition of the expectation value
\be \label{lagcdotsrag}
\lag\!\lag \cdots \rag\!\rag = \sr{\det \f{\cM_n}{2\pi}}\int \rD \cQ \ep^{-\f12 \cQ^T \cM_n \cQ} \cdots.
\ee
It is understood that $\cR_{n+1}=\cR_1$.
The wave functions $f_K(\cR_a,\cS_a)$ and $f_K^*(\cR_{a+1},\cS_a)$ are just (\ref{fKQ}) depending on the arguments
\bea
&& u_{a,k} = \sr{\f{2\ve_k}{L}} \sum_{j=1}^L \ep^{-\f{2\pi\ii j k}{L}} q_{a,j}, \nn\\
&& v_{a,k} = \sr{\f{2\ve_k}{L}} \Big( \sum_{j=1}^\ell \ep^{\f{2\pi\ii j k}{L}} q_{a+1,j}
                                    + \sum_{j=\ell+1}^L \ep^{\f{2\pi\ii j k}{L}} q_{a,j} \Big).
\eea
More explicitly, there are building bricks of the excited state wave function
\bea
&& f_k^{r,s}(\cR_a,\cS_a) = \f{1}{\sr{2^{r+s}r!s!}} H_{r,s} \Big( \f{u_{a,k}}{\sr2}, \f{u_{a,-k}}{\sr2} \Big), \nn\\
&& [ f_k^{r,s}(\cR_{a+1},\cS_a) ]^* = \f{1}{\sr{2^{r+s}r!s!}} H_{r,s} \Big( \f{v_{a,k}}{\sr2}, \f{v_{a,-k}}{\sr2} \Big).
\eea
The expectation value (\ref{lagcdotsrag}) can be evaluated by the bosonic Wick contraction following
\be
\wick{\c q_{a_1,j_1} \c q_{a_2,j_2}} = [\cM^{-1}_n]_{(a_1-1)L+j_1,(a_2-1)L+j_2}.
\ee

For $u_{a,k}$ and $v_{a,k}$, we define the vectors $U_{a,k}$, $V_{a,k}$.
Each of the vectors has $nL$ components, and the nonvanishing entries of these vectors are
\bea
&& [ U_{a,k} ]_{(a-1)L+j} = \sr{\f{2\ve_k}{L}} \ep^{-\f{2\pi\ii j k}{L}}, ~ a=1,\cdots,n, ~ j=1,\cdots,L, \nn\\
&& [ V_{a,k} ]_{(a-1)L+j} = \sr{\f{2\ve_k}{L}} \ep^{\f{2\pi\ii j k}{L}}, ~ a=1,\cdots,n, ~ j=\ell+1,\cdots,L, \nn\\
&& [ V_{a,k} ]_{a L+j} = \sr{\f{2\ve_k}{L}} \ep^{\f{2\pi\ii j k}{L}}, ~ a=1,\cdots,n-1, ~ j=1,\cdots,\ell, \nn\\
&& [ V_{n,k} ]_{j} = \sr{\f{2\ve_k}{L}} \ep^{\f{2\pi\ii j k}{L}}, ~ j=1,\cdots,\ell.
\eea
It is convenient to use the Wick contractions
\bea
&& \wick{\c u_{a_1,k_1} \c u_{a_2,k_2}} = U_{a_1,k_1}^T \cM^{-1}_n U_{a_2,k_2} , \nn\\
&& \wick{\c u_{a_1,k_1} \c v_{a_2,k_2}} = U_{a_1,k_1}^T \cM^{-1}_n V_{a_2,k_2} , \nn\\
&& \wick{\c v_{a_1,k_1} \c v_{a_2,k_2}} = V_{a_1,k_1}^T \cM^{-1}_n V_{a_2,k_2}.
\eea
It is easy to generalize (\ref{trArAKn}) to the product of the RDMs in different states
\be \label{ftrArAK1rAK2cdotsrAKntrArAGn}
  \f{\tr_A ( \r_{A,K_1} \r_{A,K_2} \cdots \r_{A,K_n} )}{\tr_A \r_{A,G}^n }
= \Big\lag\!\!\Big\lag \prod_{a=1}^n [ f_{K_a}(\cR_a,\cS_a) f_{K_a}^*(\cR_{a+1},\cS_a) ] \Big\rag\!\!\Big\rag.
\ee
The above formula generalizes the equations presented in \cite{Castro-Alvaredo:2018bij} to also the cases that a) real modes are excited and b) pairs of complex modes are excited.

\begin{figure}[p]
  \centering
  \includegraphics[height=0.9\textwidth]{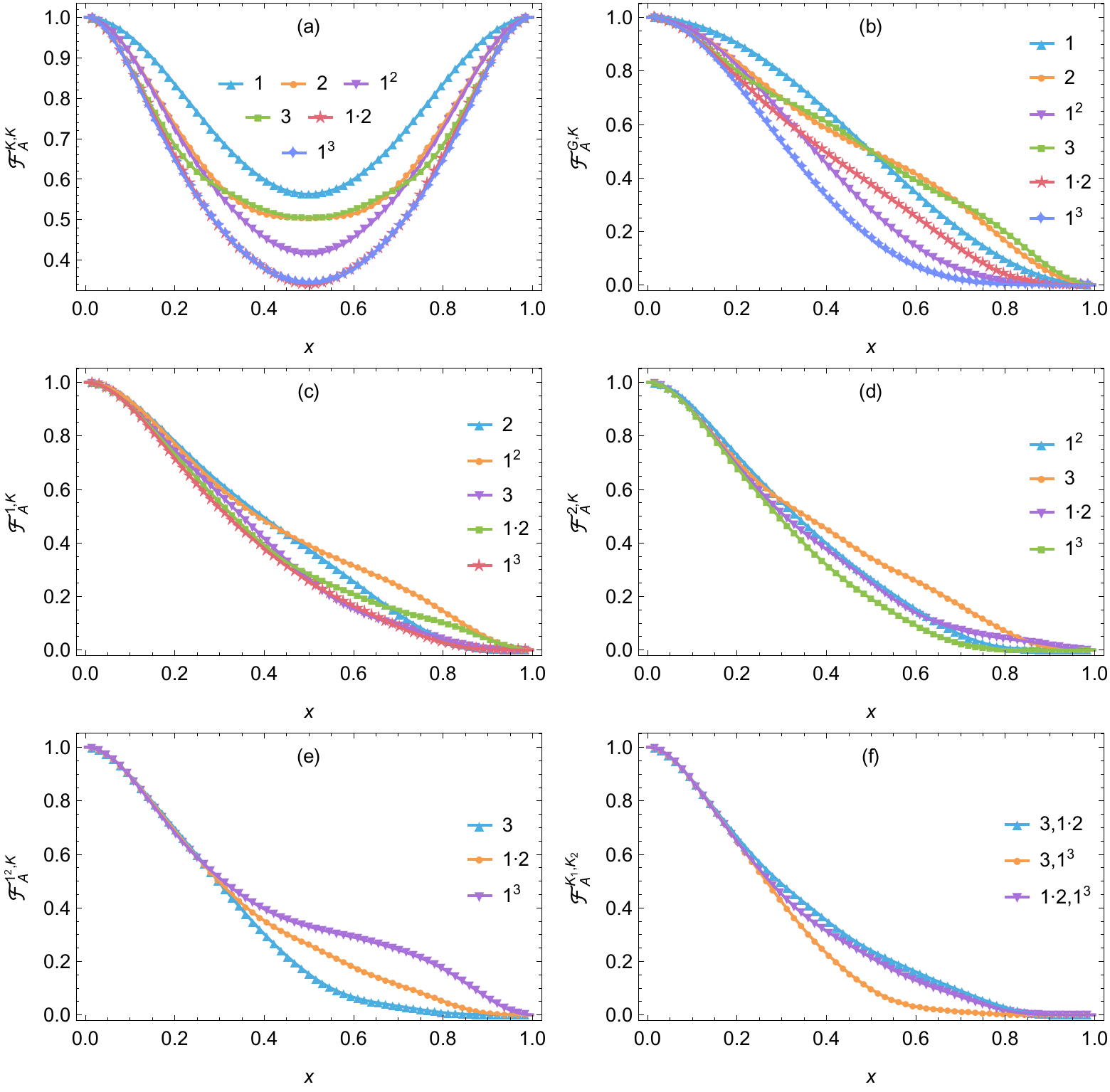}\\
  \caption{
  The product $\cF_A^{\cX_1,\cX_2}$ (\ref{FAcX1cX2CFT}) for two holomorphic operators $\cX_1$ and $\cX_2$ in the 2D free massless non-compact bosonic theory (lines) and the corresponding quantity $\cF_A^{K_1,K_2}$ (\ref{FAK1K2HC}) in the gapless limit of the harmonic chain (symbols).
  We have set $m=10^{-5}$, $L=64$.}
  \label{FAX1X2}
\end{figure}

In figure~\ref{FAX1X2}, we compare the product $\cF_A^{\cX_1,\cX_2}$ (\ref{FAcX1cX2CFT}) for two holomorphic operators $\cX_1$ and $\cX_2$ in the field theory and the corresponding quantity in the harmonic chain in the gapless limit, i.e.
\be \label{FAK1K2HC}
\cF_A^{K_1,K_2} = \f{\tr_A ( \r_{A,K_1} \r_{A,K_2})}{\tr_A \r_{A,G}^2}.
\ee
There are perfect matches between the analytical field theory results and numerical harmonic chain results.

\section{R\'enyi entropy}\label{secRE}

In this section we compare the analytical R\'enyi entropy of the excited states of the 2D free massless non-compact bosonic field theory and the corresponding numerical results in the gapless limit of the locally coupled harmonic chain.
For several low-lying states, we get the exact results of the second, the third and the fourth R\'enyi entropies, and for the general excited states we get the short interval expansion of the R\'enyi entropy with generic index $n$ and the entanglement entropy.
We also make some preliminary investigations of the excited state R\'enyi entropy in the extremely gapped harmonic chain.

\subsection{Exact results}

\begin{figure}[p]
  \centering
  \includegraphics[height=0.9\textwidth]{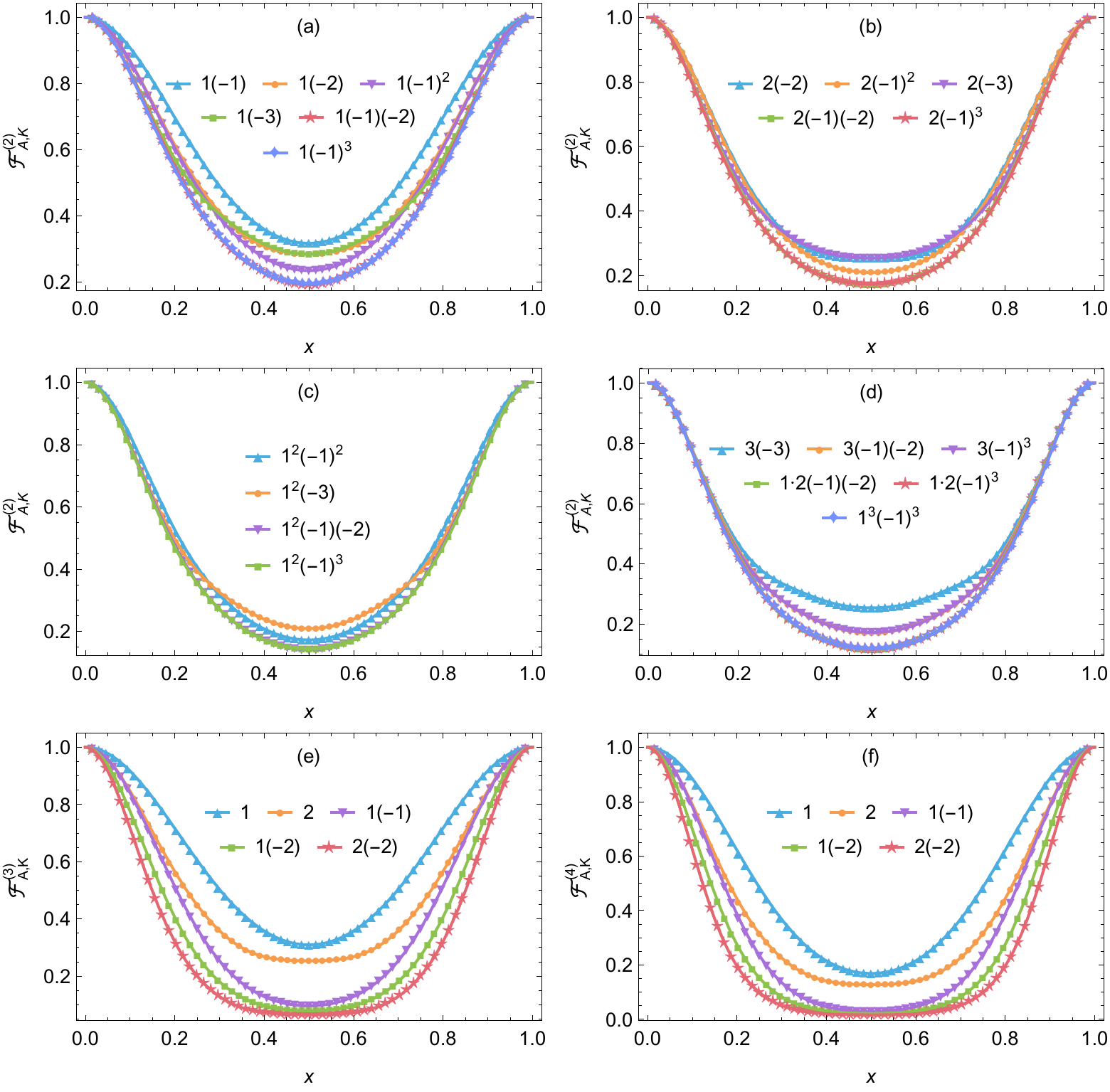}\\
  \caption{
  The second, the third and the fourth R\'enyi entropies in the excited states of the 2D free massless non-compact bosonic theory (lines) and the gapless limit of the harmonic chain (symbols).
  We have set $m=10^{-5}$, $L=64$.}
  \label{FAK234}
\end{figure}

The product (\ref{FAcX1cX2CFT}) is the ingredient to calculate the second R\'enyi entropy.
The second R\'enyi entropy $\cF_{A,\cX}^{(2)}=\cF_A^{\cX,\cX}$ in the state $|\cX\rag$ generated by the holomorphic operator $\cX$ has been calculated in subsection~\ref{subsecPro} and checked in the harmonic chain in the first panel of figure~\ref{FAX1X2}.
For a homomorphic operator $\cX$, there is a corresponding anti-holomorphic operator $\bar\cX$.
The second R\'enyi entropy $\cF_{A,\bar\cX}^{(2)}=\cF_A^{\bar\cX,\bar\cX}=\cF_A^{\cX,\cX}$ in the anti-holomorphic states $|\bar\cX\rag$ is similar and we will not show the results here.
One can also construct the state $|\cX_1\bar\cX_2\rag$ from the operator $\cX_1\bar\cX_2$ that is the product of a holomorphic operator $\cX_1$ and an anti-holomorphic operator $\bar\cX_2$.
We get the second R\'enyi entropy
\be
\cF_{A,\cX_1\bar\cX_2}^{(2)} = \cF_A^{\cX_1,\cX_1} \cF_A^{\cX_2,\cX_2}.
\ee
In the CFT, the calculation is easy, but it is nontrivial in the harmonic chain.
The third and the fourth R\'enyi entropies can be similarly constructed from the product (\ref{cFAcX1cdotscXn})
\bea
&& \cF_{A,\cX}^{(3)} = \cF_A^{\cX\cX\cX}, ~~
   \cF_{A,\cX_1\bar\cX_2}^{(3)} = \cF_A^{\cX_1,\cX_1,\cX_1} \cF_A^{\cX_2,\cX_2,\cX_2}, \nn\\
&& \cF_{A,\cX}^{(4)} = \cF_A^{\cX\cX\cX\cX}, ~~
   \cF_{A,\cX_1\bar\cX_2}^{(4)} = \cF_A^{\cX_1,\cX_1,\cX_1,\cX_1} \cF_A^{\cX_2,\cX_2,\cX_2,\cX_2}.
\eea
We show the second, the third and the fourth R\'enyi entropies for the excited states in the 2D free massless non-compact bosonic theory and the gapless limit of the harmonic chain in figure~\ref{FAK234}.
We see perfect matches for all the CFT and harmonic chain results.

\subsection{Short interval expansion}

In the previous subsection we have calculated the exact R\'enyi entropy in the 2D free massless non-compact bosonic theory for several low-lying excited states following \cite{Alcaraz:2011tn,Berganza:2011mh,Taddia:2016dbm}.
In this subsection, we will use the twist operators \cite{Calabrese:2004eu,Cardy:2007mb,Calabrese:2009qy} and their OPEs \cite{Headrick:2010zt,Calabrese:2010he,Rajabpour:2011pt,Chen:2013kpa} to calculate the short interval expansion of the single-interval R\'enyi entropy in general excited states.

In 2D CFT, the OPE of twist operators was first proposed to calculate the R\'enyi entropy of two short intervals \cite{Headrick:2010zt,Calabrese:2010he,Rajabpour:2011pt,Chen:2013kpa,Chen:2013dxa,Perlmutter:2013paa,Ruggiero:2018hyl}, and then it was used to calculate the R\'enyi entropy of one single short interval \cite{Chen:2016lbu,Lin:2016dxa,He:2017vyf}.
The twist operators in 2D CFT are primary operators in the $n$-fold CFT with scaling dimension \cite{Calabrese:2004eu}
\be
\D_n = \f{c(n^2-1)}{12n}.
\ee
For one short interval $A$ with length $\ell$ in a general translationally invariant state with density matrix $\r$, from the OPE of twist operators one can get the moments of the RDM $\r_A$ \cite{Chen:2016lbu}
\be \label{trArAcXn}
\tr_A \r_{A}^n = \f{c_n}{\ell^{2\D_n}}
               \Big( 1 + \sum_{m=1}^n \sum_{\{\mO_1, \cdots, \mO_m\}}
                 \ell^{\D_{\mO_1} + \cdots + \D_{\mO_m}}
                 b_{\mO_1\cdots\mO_m}
                 \lag\mO_1\rag_\r \cdots \lag\mO_m\rag_\r \Big),
\ee
where $\{\mO_1, \cdots, \mO_m\}$ is a collection of orthogonal nonidentity quasiprimary operators with the scaling dimensions $\{\D_{\mO_1}, \cdots, \D_{\mO_m}\}$.
The coefficients $b_{\mO_1\cdots\mO_m}$ are defined in terms of the OPE coefficients of the twist operators, see details in \cite{Chen:2016lbu}.
There are also the expectation values $\{ \lag\mO_1\rag_\r,\cdots,\lag\mO_m\rag_\r \} = \{ \tr(\r\mO_1),\cdots,\tr(\r\mO_m)\}$.

The holomorphic and anti-holomorphic sectors in the 2D massless non-compact bosonic field theory factorize, and this allows us to write (\ref{trArAcXn}) as
\bea
&& \tr_A \r_{A}^n = \f{c_n}{\ell^{2\D_n}}
               \Big( 1 + \sum_{m=1}^n \sum_{\{\mO_1, \cdots, \mO_m\}}
                 \ell^{h_{\mO_1} + \cdots + h_{\mO_m}}
                 b_{\mO_1\cdots\mO_m}
                 \lag\mO_1\rag_\r \cdots \lag\mO_m\rag_\r \Big) \nn\\
&& \phantom{\tr_A \r_{A}^n =} \times
               \Big( 1 + \sum_{m=1}^n \sum_{\{\bar\mO_1, \cdots, \bar\mO_m\}}
                   \ell^{\bar h_{\bar \mO_1} + \cdots + \bar h_{\bar \mO_m}}
                   b_{\bar\mO_1\cdots\bar\mO_m}
                   \lag\bar\mO_1\rag_\r \cdots \lag\bar\mO_m\rag_\r \Big),
\eea
where $\{\mO_1, \cdots, \mO_m\}$ is a collection of purely holomorphic orthogonal nonidentity quasiprimary operators with the conformal weights $\{h_{\mO_1}, \cdots, h_{\mO_m}\}$, and $\{\bar\mO_1, \cdots, \bar\mO_m\}$ is the set of purely anti-holomorphic orthogonal nonidentity quasiprimary operators with the conformal weights $\{\bar h_{\bar\mO_1}, \cdots, \bar h_{\bar \mO_m}\}$.

In this paper, we only consider the states $\r_\cX$ with the vanishing expectation values $\lag J \rag_\cX=\lag \bar J \rag_\cX=\lag \cK \rag_\cX=\lag \bar\cK \rag_\cX=0$.
Then we get the R\'enyi entropy
\bea \label{cFAcXn}
&& \cF_{A,\cX}^{(n)} = \big\{ 1
   + \ell^2 b_T ( \lag T \rag_\cX - \lag T \rag_G )
   + \ell^4  [ b_{TT}  ( \lag T \rag_\cX^2 - \lag T \rag_G^2  )
                - b_T^2 \lag T \rag_G  ( \lag T \rag_\cX - \lag T \rag_G  ) \nn\\
&& \phantom{\cF_{A,\cX}^{(n)} =}
                + b_\cA  ( \lag \cA \rag_\cX - \lag \cA \rag_G  )  ] + o(\ell^4)  \big\}
 \big\{ 1
   + \ell^2 b_{\bar T} ( \lag \bar T \rag_\cX - \lag \bar T \rag_G )
   + \ell^4  [ b_{\bar T \bar T}  ( \lag \bar T \rag_\cX^2 - \lag \bar T \rag_G^2  )  \nn\\
&& \phantom{\cF_{A,\cX}^{(n)} =}
                - b_{\bar T}^2 \lag \bar T \rag_G  ( \lag \bar T \rag_\cX - \lag \bar T \rag_G  )
                + b_{\bar \cA}  ( \lag \bar \cA \rag_\cX - \lag \bar \cA \rag_G  )  ] + o(\ell^4)  \big\},
\eea
which is a product of the holomorphic and anti-holomorphic parts.
There are simple relations $b_{\bar T}=b_T$, $b_{\bar \cA}=b_\cA$ and $b_{\bar T\bar T}=b_{TT}$, and the coefficients $b_T$, $b_{\cA}$ and $b_{TT}$ can be found in \cite{Chen:2016lbu,Lin:2016dxa,He:2017vyf}
\be \label{bTbTTbcA}
b_T = \frac{n^2-1}{12 n}, ~~
b_\cA = \frac{(n^2-1)^2}{288 n^3}, ~~
b_{TT} = \frac{(n^2-1) (5 n^3-3 n^2-5 n+27)}{1440 n^3}.
\ee
The operator $\cB$ with conformal weight $h_\cB=4$, which is defined in (\ref{cAcBdef}), is primary with the vanishing coefficient $b_\cB=0$, and so $\cB$ and its anti-holomorphic counterpart $\bar\cB$ do not contribute to the R\'enyi entropy up to the order $\ell^4$.

Especially, at the leading nontrivial order $\ell^2$, we have
\be
\cF_{A,\cX}^{(n)} = 1 + \ell^2 b_T [( \lag T\rag_\mX - \lag T\rag_G ) + ( \lag\bar T\rag_\mX - \lag\bar T\rag_G ) ] + o(\ell^2),
\ee
with the expectation values
\be \label{TTbevs}
\lag T \rag_G = \lag \bar T \rag_G = \frac{\pi^2 c }{6 L^2}, ~~
\lag T \rag_\cX = \frac{\pi^2 (c-24 h_\cX)}{6 L^2}, ~~
\lag \bar T \rag_\cX = \frac{\pi^2 (c-24 \bar h_\cX)}{6 L^2}.
\ee
Note that here the central charge is $c=1$.
Then we obtain the leading order R\'enyi entropy in short interval expansion
\be \label{RenyiLeadingUniversal}
\cF_{A,\cX}^{(n)} = 1 - \f{\pi^2(n^2-1)\D_\cX x^2}{3n} + o(x^2).
\ee
The scaling dimension in the 2D free massless non-compact bosonic theory has the following correspondence in the gapless harmonic chain
\be \label{Kabsdef}
\D_\cX=h_\cX+\bar h_\cX \lra |K| = \sum_k r_k |k|.
\ee

Taking $n\to1$ limit, we get the difference of the excited state entanglement entropy from that of the ground state
\be
\d S_{A,\cX} = \ell^2 a_T [( \lag T\rag_\mX - \lag T\rag_G ) + ( \lag\bar T\rag_\mX - \lag\bar T\rag_G ) ]
             + \ell^4 a_{TT} [( \lag T\rag_\mX^2 - \lag T\rag_G^2 ) + ( \lag\bar T\rag_\mX^2 - \lag\bar T\rag_G^2 ) ]
             + o(\ell^4),
\ee
with the coefficients \cite{He:2017txy}
\be
a_T = -\frac{1}{6}, ~~
a_{TT} = -\frac{1}{30}.
\ee
Using the expectation values of the stress tensor (\ref{TTbevs}), we further get the entanglement entropy
\be
\d S_{A,\cX} =
  \frac{2 \pi^2(h_\cX+ \bar h_\cX) x^2 }{3}
- \frac{2 \pi^4( 12 h_\cX^2 - h_\cX + 12 \bar h_\cX^2 - \bar h_\cX ) x^4 }{45}
+ o(x^4).
\ee
This is consistent with the special cases of the exact entanglement entropies \cite{Essler2013ShellFilling,Calabrese:2014Entanglement}
\be
\d S_{A,J} = \d S_{A,\bar J} =  \f12 \d S_{A,J\bar J} =
- 2 \Big[ \sin \f{\pi\ell}{L}
                            + \log\Big( 2 \sin \f{\pi\ell}{L} \Big)
                            + \psi\Big(\f12 \csc\f{\pi\ell}{L}\Big) \Big],
\ee
with $\psi$ denoting the digamma function. Note that up to order $x^2$ the states with larger conformal weights have larger entanglement entropies but this pattern can change for larger $x$, see figure~\ref{FAX1X2} and figure~\ref{FAK234}.

To evaluate the R\'enyi entropy at order $\ell^4$, we need to know the expectation values $\lag \cA \rag_\cX$, $\lag \bar \cA \rag_\cX$.
Under a general conformal transformation $z \to f(z)$, the quasiprimary operator $\cA$ transforms as
\be
\cA(z) = f'^4 \cA(f) + \f{9}{10} s f'^2 T(f) + \f{3}{80}s^2,
\ee
with the Schwarzian derivative $s(z)$ (\ref{Schwarzian}).
In the holomorphic states
\be
K =
\{
\{G\},
\{1\},
\{2,1^2\},
\{3,1\!\cdot\!2,1^3\},
\{4,1\!\cdot\!3,2^2,1^22,1^4\},
\{5,1\!\cdot\!4,2\!\cdot\!3,1^23,12^2,1^32,1^5\}
\},
\ee
we get the respective expectation values
\bea
&& \lag \cA \rag_K =
\Big\{
\Big\{\frac{3 \pi^4}{20 L^4}\Big\},
\Big\{\frac{243 \pi^4}{20 L^4}\Big\},
\Big\{\frac{2403 \pi^4}{20 L^4},\frac{1443 \pi^4}{20 L^4}\Big\},
\Big\{\frac{8403 \pi^4}{20 L^4},\frac{6483 \pi^4}{20 L^4},\frac{3603 \pi^4}{20 L^4}\Big\}, \nn\\
&& \phantom{\lag \cA \rag_K =}
\Big\{\frac{20163 \pi^4}{20 L^4},\frac{14403 \pi^4}{20 L^4},\frac{8643 \pi^4}{20 L^4},\frac{11523 \pi^4}{20 L^4},\frac{6723 \pi^4}{20 L^4}\Big\},  \nn\\
&& \phantom{\lag \cA \rag_K =}
\Big\{\frac{39603 \pi^4}{20 L^4},\frac{28083 \pi^4}{20 L^4},\frac{22323 \pi^4}{20 L^4},\frac{21363 \pi^4}{20 L^4},\frac{16563 \pi^4}{20 L^4},\frac{17523 \pi^4}{20 L^4},\frac{10803 \pi^4}{20 L^4}\Big\} \Big\}.
\eea

\begin{figure}[p]
  \centering
  \includegraphics[height=1.17\textwidth]{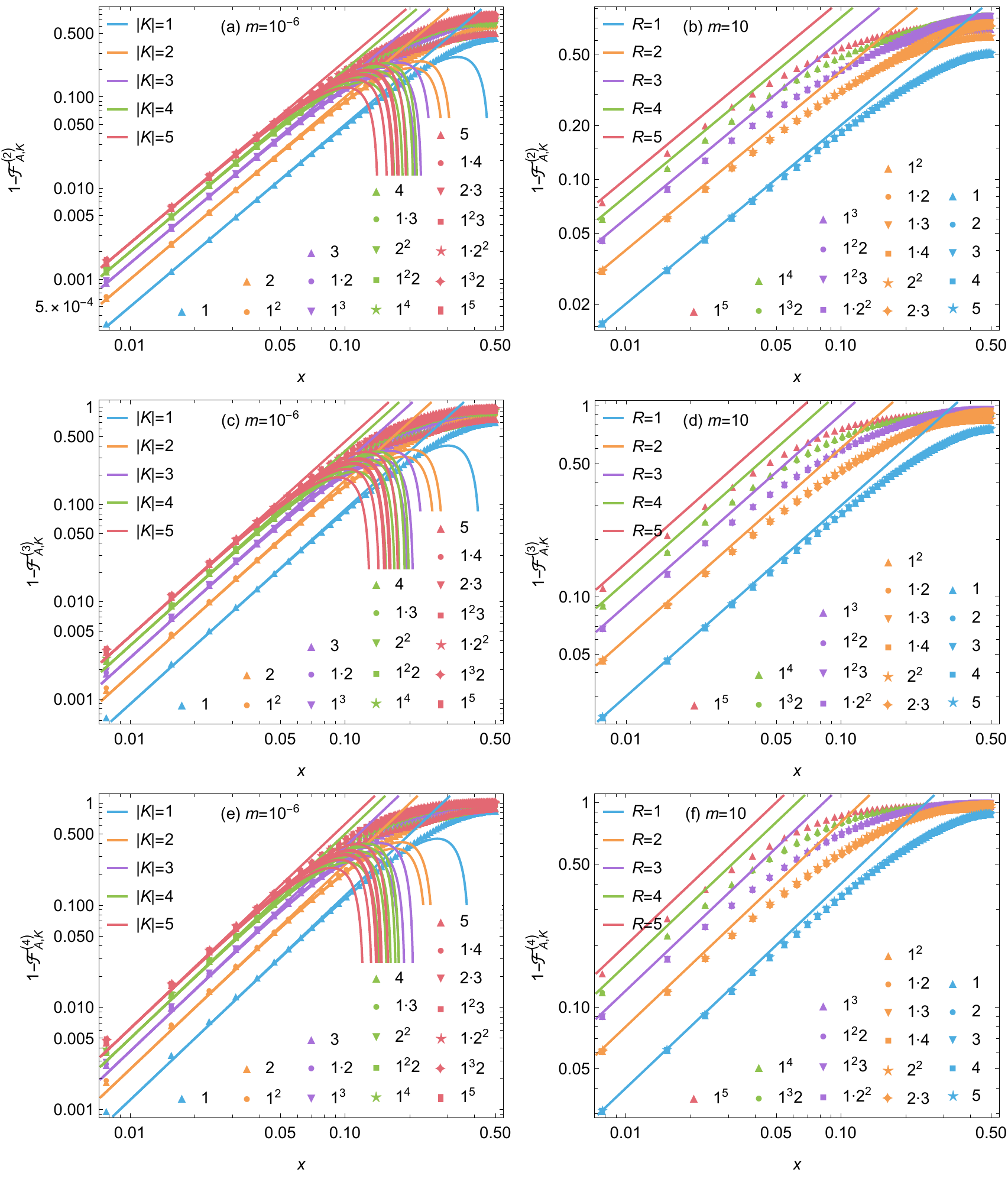}\\
  \caption{The short interval expansion of the excited state R\'enyi entropy in the massless (left column) and extremely massive (right column) limits of the 2D free non-compact bosonic theory (lines) and the excited state R\'enyi entropy in the gapless (left column) and extremely gapped (right column) limits of the harmonic chain (symbols).
  In the left column, we show not only the short interval expansion results to the leading nontrivial order $x^2$ (straight lines), but also the results to the next-to-leading order $x^4$ (curved lines).
  We have set $L=128$.}
  \label{RenyiGaplessGapped}
\end{figure}

We compare the short interval expansion of the excited state R\'enyi entropy in the 2D free massless non-compact bosonic theory and the excited state R\'enyi entropy in the gapless limit of the harmonic chain in the left panels of figure~\ref{RenyiGaplessGapped}.
In the figure we only show the special cases of states $\r_\cX$ with the vanishing anti-holomorphic conformal weights $\bar h_\cX=0$.
We find good matches for small $x=\f{\ell}{L}$.

It is interesting to compare how the R\'enyi entropy changes among different states in different models.
In the extremely massive limit of the free bosonic theory, i.e. the extremely gapped limit of the harmonic chain, the R\'enyi entropy in the excited state that both the quasiparticle momenta and the momentum differences are large takes the universal form \cite{Castro-Alvaredo:2018dja,Castro-Alvaredo:2018bij,Castro-Alvaredo:2019irt,Castro-Alvaredo:2019lmj}
\be \label{cFAKnuniv}
\cF_{A,K}^{(n)} = \prod_k \Big\{ \sum_{p=0}^{r_k} [ C_{r_k}^p x^p (1-x)^{r_k-p} ]^n \Big\}.
\ee
Generally when the momentum differences of the excited quasiparticles are small, there exist additional contributions to the R\'enyi entropy \cite{Zhang:2020vtc,Zhang:2020dtd}, however, in this paper we only consider the leading term in the small interval expansion, a circumstance in which  the universal R\'enyi entropy (\ref{cFAKnuniv}) still applies.
We obtain
\be \label{RenyiLeadingUniversal2}
\cF_{A,K}^{(n)} = 1 - n R x + o(x),
\ee
with the total number of the excited quasiparticles
\be \label{Rdef}
R = \sum_k r_k.
\ee
The to the best of our knowledge the above equation has not been discussed in the literature previously. We compare the analytical short interval expansion of the R\'enyi entropy and the excited state R\'enyi entropy in the extremely gapped limit of the harmonic chain in the right panels of figure~\ref{RenyiGaplessGapped}.
There are good matches for small $x=\f{\ell}{L}$.
We report more details on the universal R\'enyi entropy and its corrections in the gapped harmonic chain in \cite{Zhang:2020vtc,Zhang:2020dtd}.

\section{Schatten distance}\label{secSD}

In this section we compare the analytical results of the second and the fourth subsystem Schatten distances among the ground and excited states in the 2D free massless non-compact bosonic field theory and the corresponding numerical results in the harmonic chain with local interactions.
We also mention some preliminary results of the Schatten distance in the extremely gapped harmonic chain.

\subsection{Exact results}

When the second R\'enyi entropies $\cF_A^{\cX_1,\cX_1}$ and $\cF_A^{\cX_2,\cX_2}$ of the two RDMs $\r_{A,\cX_1}$ and $\r_{A,\cX_2}$ are known, we only need the product $\cF_A^{\cX_1,\cX_2}$ to calculate the second Schatten distance
\be
D_2(\r_{A,\cX_1},\r_{A,\cX_2}) = \Big[ \f12 \big( \cF_A^{\cX_1,\cX_1} - 2 \cF_A^{\cX_1,\cX_2} + \cF_A^{\cX_2,\cX_2} \big) \Big]^{1/2}.
\ee
Between the ground state RDM $\r_{A,G}$ and the holomorphic excited state RDM $\r_{A,\cX}$ and between two different holomorphic excited state RDMs $\r_{A,\cX_1}$ and $\r_{A,\cX_2}$, the products $\cF_A^{G,\cX}$ and $\cF_A^{\cX_1,\cX_2}$ have been obtained in section~\ref{secCFT} and compared to their harmonic chain counterparts in figure~\ref{FAX1X2}.

\begin{figure}[p]
  \centering
  \includegraphics[height=1.2\textwidth]{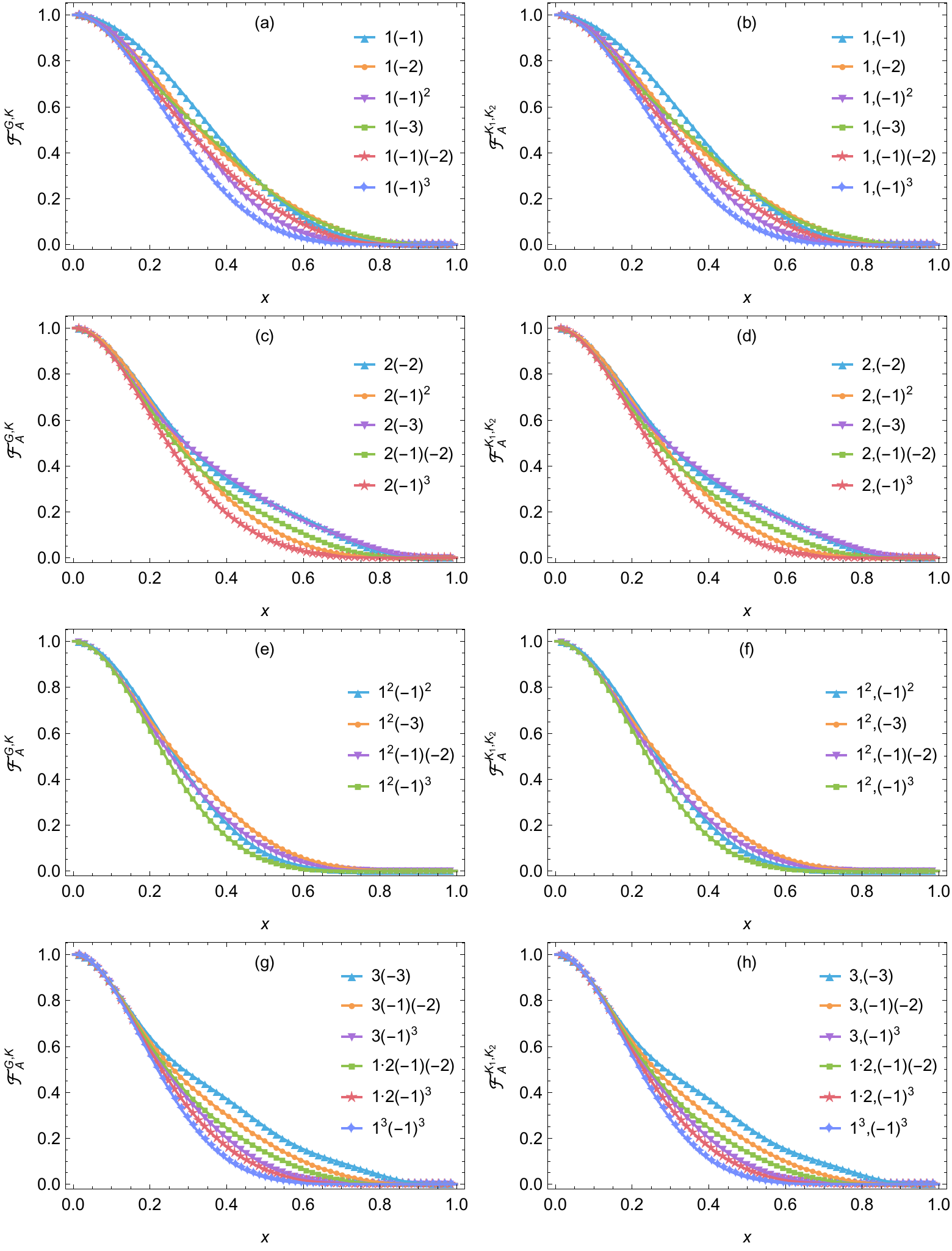}\\
  \caption{The product $\cF_A^{G,\cX_1\bar\cX_2}$ (left column) and $\cF_A^{\cX_1,\bar\cX_2}$ (right column) in the 2D free massless non-compact bosonic theory (lines) and the corresponding quantities in the gapless limit of the harmonic chain (symbols). We have set $m=10^{-5}$, $L=64$.}
  \label{FAGX1X2bFAX1X2b}
\end{figure}

\begin{figure}[p]
  \centering
  \includegraphics[height=0.9\textwidth]{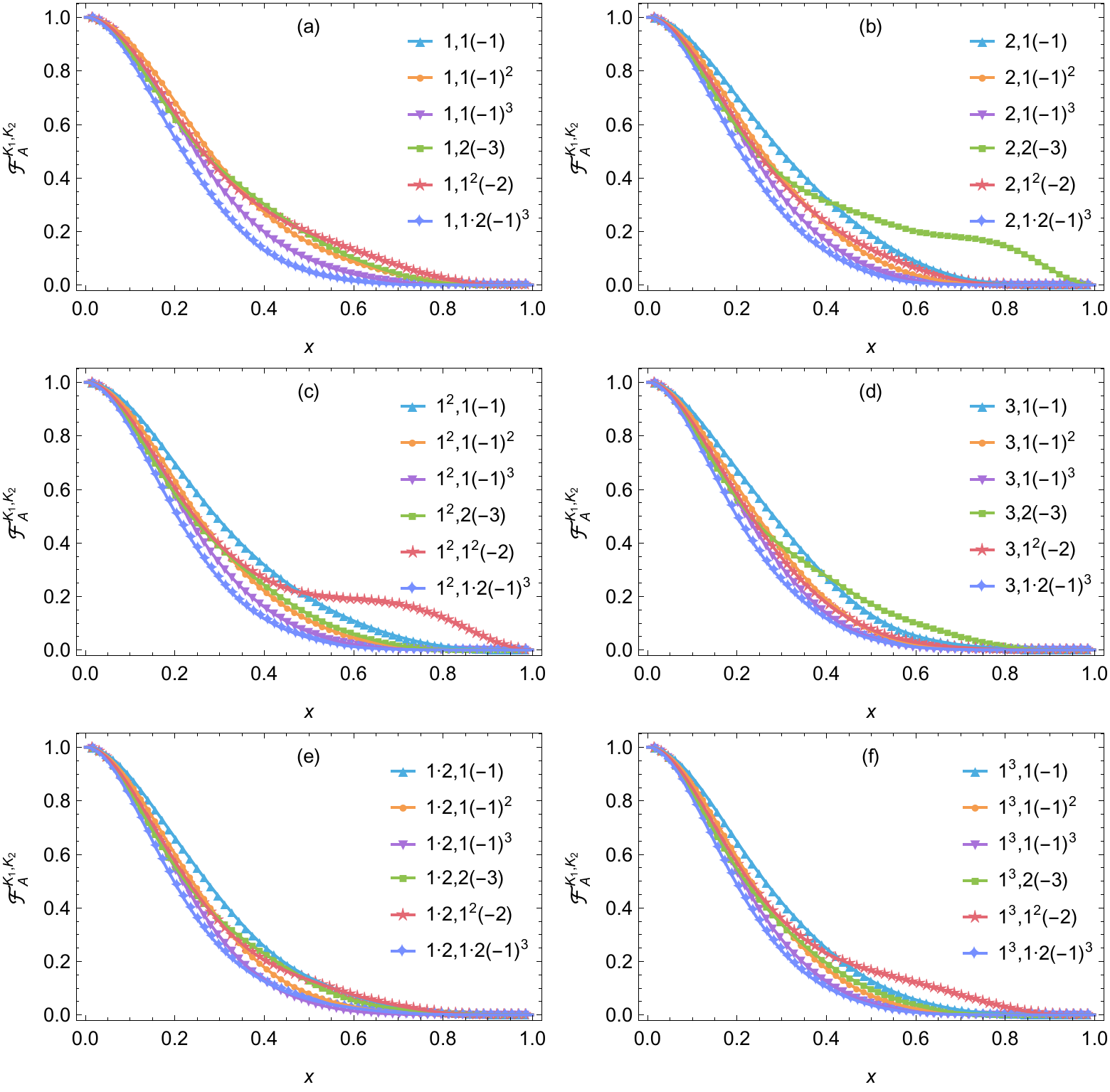}\\
  \caption{The product $\cF_A^{\cX_1,\cX_2\bar\cX_3}$ in the 2D free massless non-compact bosonic theory (lines) and the corresponding quantities in the gapless limit of the harmonic chain (symbols). We have set $m=10^{-5}$, $L=64$.}
  \label{FAX1X2X3b}
\end{figure}

\begin{figure}[p]
  \centering
  \includegraphics[height=0.9\textwidth]{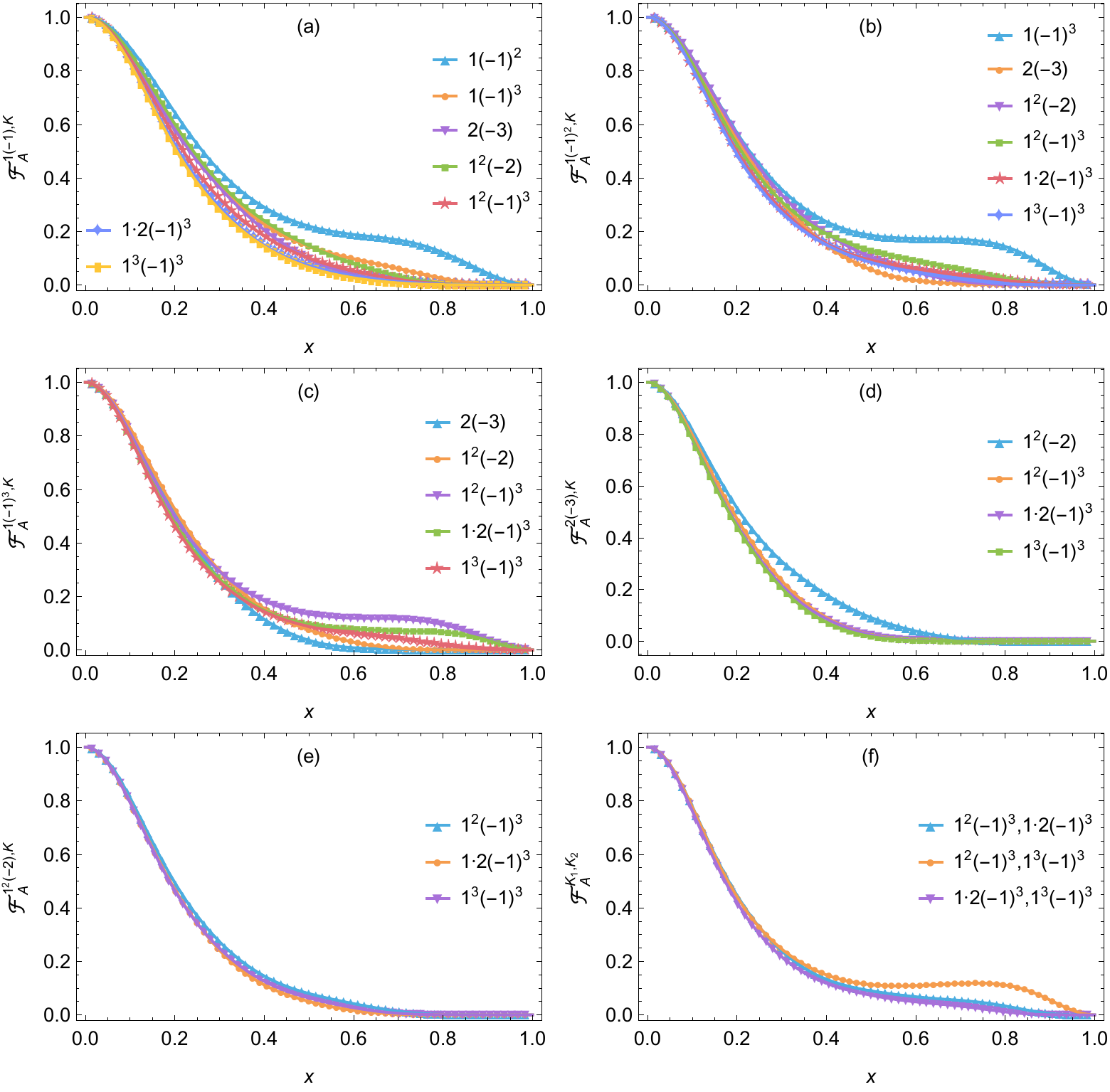}\\
  \caption{The product $\cF_A^{\cX_1\bar\cX_2,\cX_3\bar\cX_4}$ in the 2D free massless non-compact bosonic theory (lines) and the corresponding quantities in the gapless limit of the harmonic chain (symbols). We have set $m=10^{-5}$, $L=64$.}
  \label{FAX1X2bX3X4b}
\end{figure}

\begin{figure}[tp]
  \centering
  \includegraphics[height=0.6\textwidth]{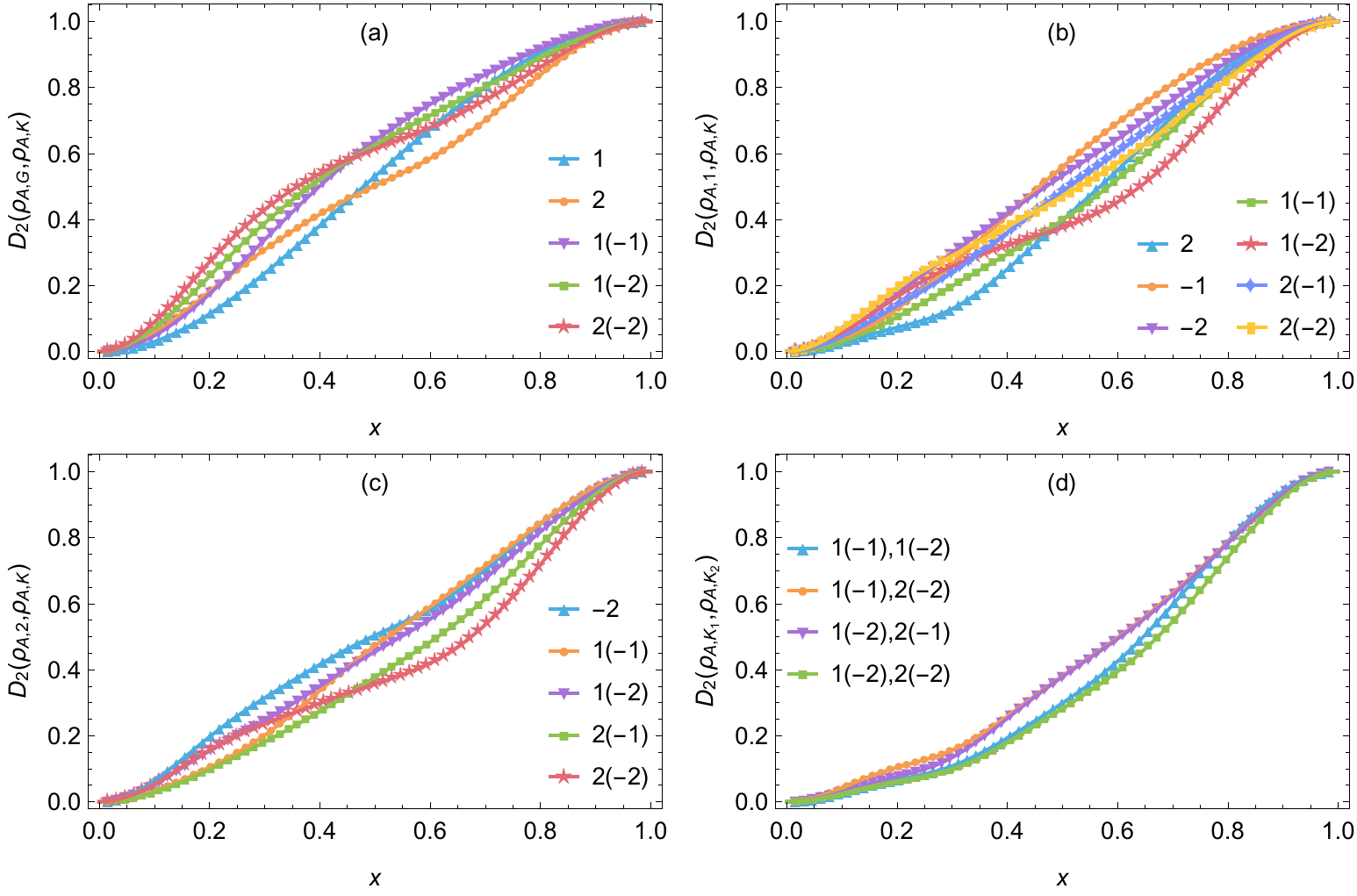}\\
  \caption{The second Schatten distance in the 2D free massless non-compact bosonic field theory (lines) and the gapless limit of the  harmonic chain (symbols). We have set $m=10^{-5}$, $L=64$.}
  \label{D2AK1K2}
\end{figure}

For the RDMs of the states excited by the holomorphic operator $\cX_1$, the anti-holomorphic operator $\bar\cX_2$ and the product operator $\cX_1\bar\cX_2$, we get the equal products
\be
\cF_A^{G,\cX_1\bar\cX_2} = \cF_A^{\cX_1,\bar\cX_2} = \cF_A^{G,\cX_1} \cF_A^{G,\cX_2}.
\ee
This is of no surprise in the CFT but it is nontrivial in the harmonic chain.
We check this in figure~\ref{FAGX1X2bFAX1X2b}.
Between the RDMs of the holomorphic state $|\cX_1\rag$ and the RDMs in the states of the product operators $|\cX_2\bar\cX_3\rag$, we get
\be
\cF_A^{\cX_1,\cX_2\bar\cX_3} = \cF_A^{\cX_1,\cX_2} \cF_A^{G,\cX_3}.
\ee
This is checked in figure~\ref{FAX1X2X3b}.
Between the RDMs in the states of the product operators $|\cX_1\bar\cX_2\rag$ and $|\cX_3\bar\cX_4\rag$, we get
\be
\cF_A^{\cX_1\bar\cX_2,\cX_3\bar\cX_4} = \cF_A^{\cX_1,\cX_3} \cF_A^{\cX_2,\cX_4}.
\ee
This is checked in figure~\ref{FAX1X2bX3X4b}.
 In figure~\ref{D2AK1K2} we provide some examples of the second Schatten distance in the 2D free massless non-compact bosonic field theory and the gapless limit of the harmonic chain.

\begin{figure}[tp]
  \centering
  \includegraphics[height=0.6\textwidth]{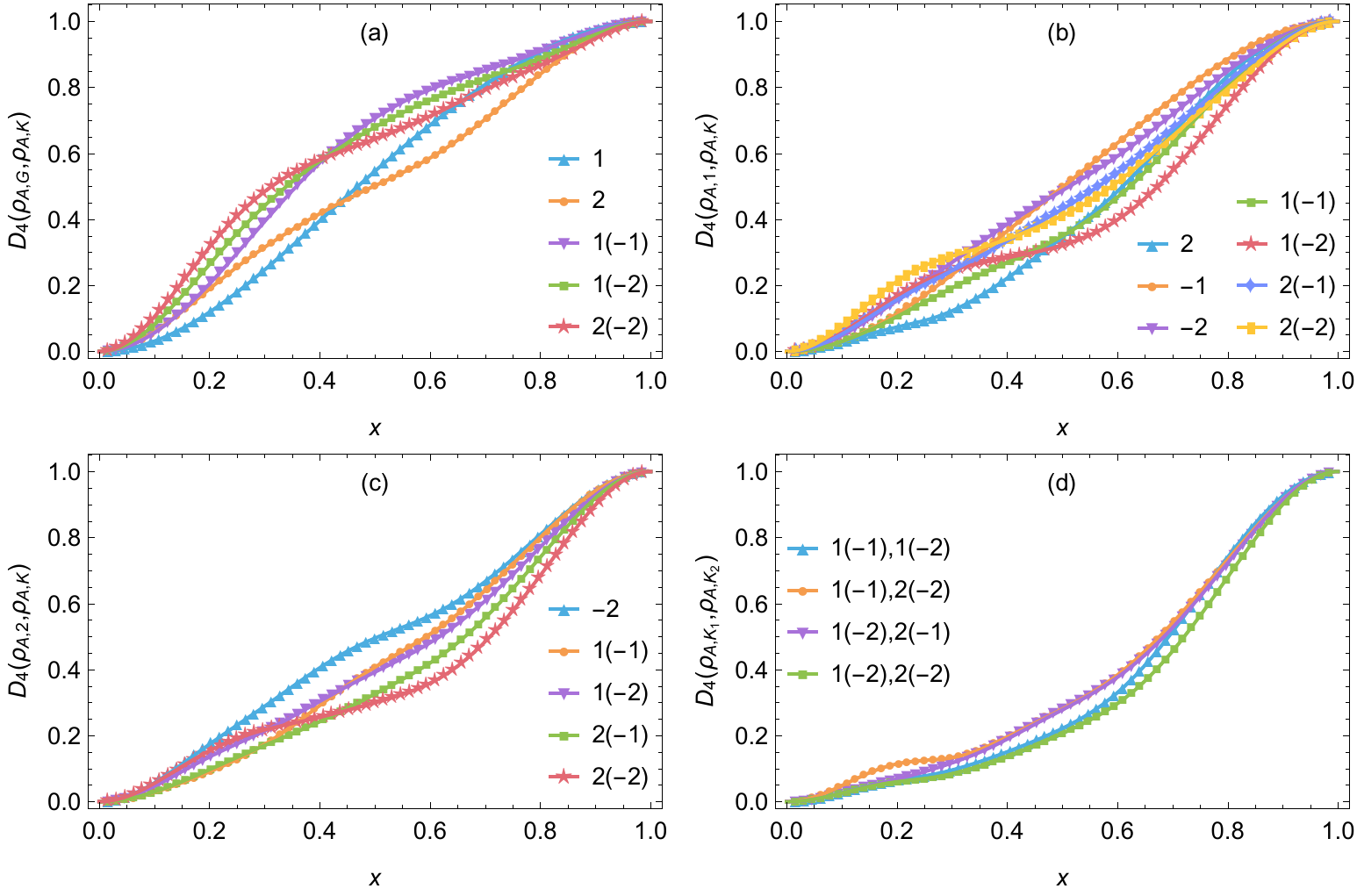}\\
  \caption{The fourth Schatten distance in the 2D free massless non-compact bosonic field theory (lines) and the gapless limit of the  harmonic chain (symbols). We have set $m=10^{-5}$, $L=64$.}
  \label{D4AK1K2}
\end{figure}

From the product (\ref{cFAcX1cdotscXn}) with $n=4$, we get the fourth Schatten distance
\bea
&& D_4( \r_{A,\cX_1\bar\cX_2} , \r_{A,\cX_3\bar\cX_4} )^4 =
   \f12 \big(
       \cF_A^{\cX_1,\cX_1,\cX_1,\cX_1} \cF_A^{\cX_2,\cX_2,\cX_2,\cX_2}
   - 4 \cF_A^{\cX_1,\cX_1,\cX_1,\cX_3} \cF_A^{\cX_2,\cX_2,\cX_2,\cX_4} \nn\\
&& \phantom{D_4( \r_{A,\cX_1\bar\cX_2} , \r_{A,\cX_3\bar\cX_4} )^4 =}
   + 4 \cF_A^{\cX_1,\cX_1,\cX_3,\cX_3} \cF_A^{\cX_2,\cX_2,\cX_4,\cX_4}
   + 2 \cF_A^{\cX_1,\cX_3,\cX_1,\cX_3} \cF_A^{\cX_2,\cX_4,\cX_2,\cX_4} \nn\\
&& \phantom{D_4( \r_{A,\cX_1\bar\cX_2} , \r_{A,\cX_3\bar\cX_4} )^4 =}
   - 4 \cF_A^{\cX_1,\cX_3,\cX_3,\cX_3} \cF_A^{\cX_2,\cX_4,\cX_4,\cX_4}
   +   \cF_A^{\cX_3,\cX_3,\cX_3,\cX_3} \cF_A^{\cX_4,\cX_4,\cX_4,\cX_4}
   \big).
\eea
We compare the results in the 2D free massless non-compact bosonic field theory and the gapless limit of the harmonic chain in figure~\ref{D4AK1K2}.

\subsection{Short interval expansion}

For two RDMs $\r_A,\s_A$ of an interval with a short length $\ell$, using the OPE of twist operators,  one can find \cite{He:2017txy}
\be
\tr_A(\r_A-\s_A)^n = \f{c_n}{\ell^{2\D_n}}
\sum_{\{\cO_1,\cdots,\cO_n\}}
\ell^{\D_{\cO_1}+\cdots+\D_{\cO_n}}
b_{\cO_1\cdots\cO_n}
\big(\lag \cO_1 \rag_\r - \lag \cO_1 \rag_\s\big)
\cdots
\big(\lag \cO_n \rag_\r - \lag \cO_n \rag_\s\big).
\ee
The coefficient $b_{\cO_1\cdots\cO_n}$ is defined the same as that in (\ref{trArAcXn}).
In two energy eigenstates $\r_{\cX_1},\r_{\cX_2}$ with the same expectation values of the current $\lag J \rag_{\cX_1}=\lag J \rag_{\cX_2}$, $\lag \bar J \rag_{\cX_1}=\lag \bar J \rag_{\cX_2}$, we obtain the leading order short interval expansion results
\bea
&& \f{\tr_A(\r_{\cX_1}-\r_{\cX_2})^2}{\tr_A\r_{A,G}^2} = \ell^4 [ b_{TT} \d \lag T \rag^2 + b_{T\bar T} \d \lag T \rag \d \lag \bar T \rag + b_{\bar T \bar T} \d \lag \bar T \rag^2 ]|_{n \to 2} + o(\ell^4), \nn\\
&& \f{\tr_A(\r_{\cX_1}-\r_{\cX_2})^4}{\tr_A\r_{A,G}^4} =
   \ell^8 [ b_{TTTT} \d \lag T \rag^4
          + b_{TTT\bar T} \d \lag T \rag^3 \d \lag \bar T \rag
          + b_{TT\bar T\bar T} \d \lag T \rag^2 \d \lag \bar T \rag^2 \nn\\
&& \phantom{\f{\tr_A(\r_{\cX_1}-\r_{\cX_2})^4}{\tr_A\r_{A,G}^4} =}
          + b_{T\bar T\bar T\bar T} \d \lag T \rag \d \lag \bar T \rag^3
          + b_{\bar T \bar T \bar T \bar T} \d \lag \bar T \rag^4 ]|_{n \to 4} + o(\ell^8).
\eea
We have the $n$-dependent coefficients
\bea
&& b_{T\bar T} = \f{n-1}{n} b_T^2, ~~
   b_{\bar T\bar T} = b_{TT}, ~~
   b_{TTT\bar T}=b_{T\bar T\bar T\bar T}=\f{n-3}{n}b_{TTT}b_T, \nn\\
&& b_{TT\bar T\bar T} = \f{(n-2)(n-3)}{n(n-1)} b_{TT}^2, ~~
   b_{\bar T\bar T\bar T\bar T}=b_{TTTT},
\eea
and one can find $b_T$, $b_{TT}$, $b_{TTT}$ and $b_{TTTT}$ in \cite{Chen:2016lbu,Lin:2016dxa,He:2017vyf}.
We have shown $b_T$ and $b_{TT}$ in (\ref{bTbTTbcA}), and there are also \cite{Chen:2016lbu,Lin:2016dxa,He:2017vyf}
\bea
&& b_{TTT} = \frac{(n^2-1)(n-2)(35 n^5+7 n^4-86 n^3+458 n^2-717 n-2001 )}{362880 n^5}, \nn\\
&& b_{TTTT} = \frac{(n^2-1)(n-2)(n-3)}{87091200 n^7} (175 n^7+245 n^6-761 n^5+4013 n^4 \nn\\
&& \phantom{b_{TTTT} =}
-11195 n^3-11545 n^2+141381 n+188727).
\eea
We have the expectation value differences
\be
\d \lag T \rag = -\f{4\pi^2\d h}{L^2}, ~~
\d \lag \bar T \rag = -\f{4\pi^2\d \bar h}{L^2},
\ee
with the differences of the conformal weights
\be
\d h = h_{\cX_1} - h_{\cX_2}, ~~
\d \bar h = \bar h_{\cX_1} - \bar h_{\cX_2}.
\ee
Then we get the short interval expansion of the Schatten distances
\bea \label{D2D4}
&& D_2( \r_{A,\cX_1}, \r_{A,\cX_2} )^2 = \frac{\pi^4 x^4}{32} ( 3 \d h^2+2 \d h \d\bar h+3 \d\bar h^2 ) + o(x^4), \nn\\
&& D_4( \r_{A,\cX_1}, \r_{A,\cX_2} )^4 = \frac{3 \pi^8 x^8}{131072} ( 467 \d h^4+1220 \d h^3 \d\bar h+1922 \d h^2 \d\bar h^2 \nn\\
&& \phantom{D_4( \r_{A,\cX_1}, \r_{A,\cX_2} )^4 =}
                                        + 1220 \d h \d\bar h^3+467 \d\bar h^4 )
                                        + o(x^8).
\eea
Note that up to the calculated order the distances are just functions of the $\d h$ and $\d \bar h$. However, this pattern does not survive for finite $x$ in general and many exotic features starts appearing. For example, in figure~\ref{D2AK1K2} the state $|1\rag$ seems closer to the ground state $|G\rag$ than the state $|2\rag$ for small $x$ but for larger $x$ it is reversed. Similar structure is also present for more complicated states too which we have no good physical explanation. The above equations and their generalizations to arbitrary $n$ might be useful for later computation of the trace distance and in the study of the thermalization after quantum quenches.

\begin{figure}[t]
  \centering
  \includegraphics[height=0.78\textwidth]{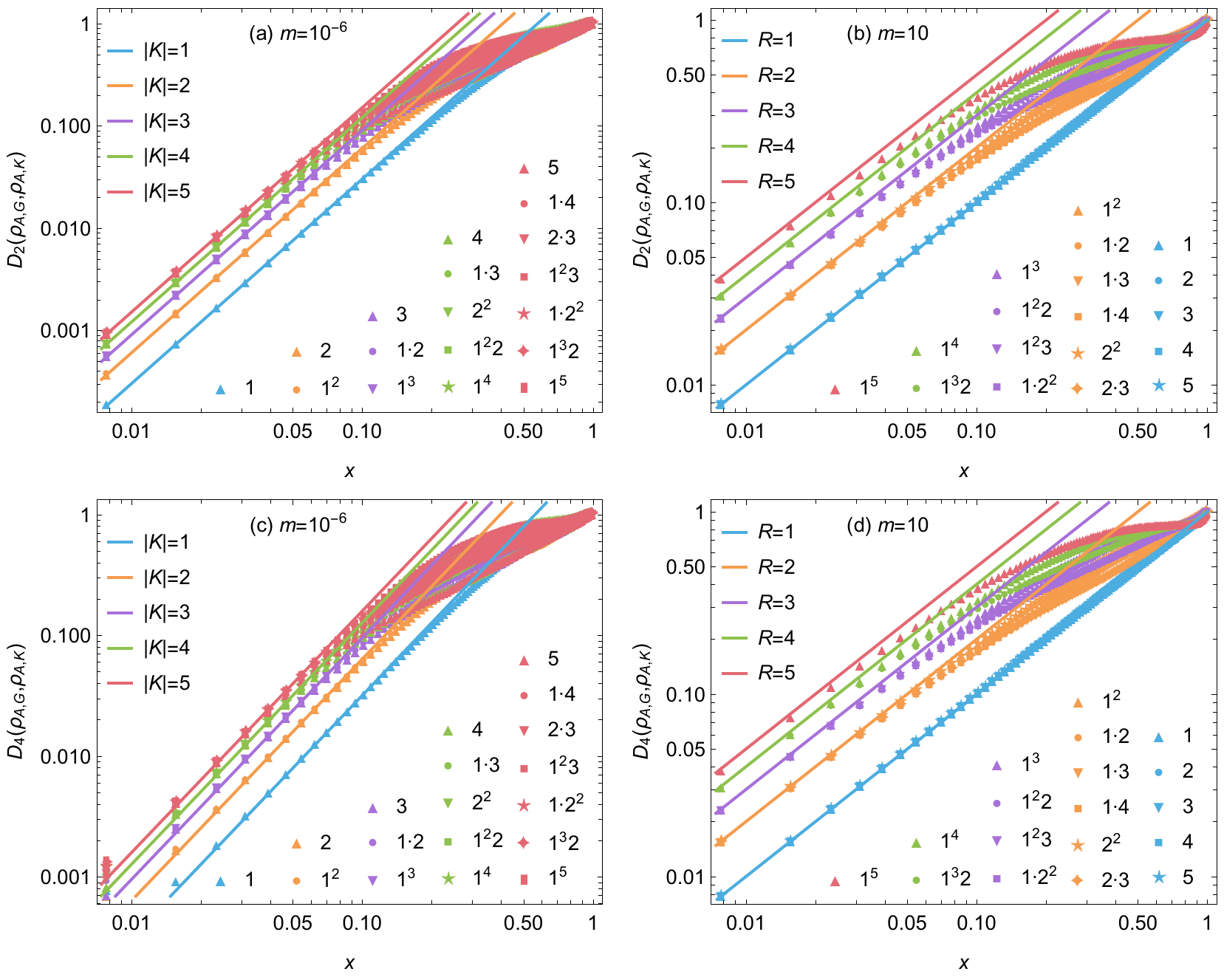}\\
  \caption{The short interval expansion of the Schatten distance in the massless (left column) and extremely massive (right column) limits of the 2D free non-compact bosonic theory (lines) and the Schatten distance in the gapless (left column) and extremely gapped (right column) limits of the harmonic chain (symbols). We have set $L=128$.}
  \label{SchattenGaplessGapped}
\end{figure}

For the special case of the holomorphic state generated by $\cX$ with $\bar h_\cX=0$, we have%
\footnote{It is easy to see that $D_n( \r_{A,G}, \r_{A,\cX} ) \pp x^{2}+o(x^2)$, but it is difficult to obtain a closed form of the overall coefficient for a general even integer $n$ as one needs a closed form of the $n$-point correlation function of the stress tensor.
One could see details in \cite{Zhang:2019itb}.}
\bea
&& D_2( \r_{A,G}, \r_{A,\cX} )^2 = \frac{3 \pi^4 h_\cX^2 x^4}{32} + o(x^4), \nn\\
&& D_4( \r_{A,G}, \r_{A,\cX} )^4 = \frac{1401\pi^8 h_\cX^4 x^8}{131072} + o(x^8).
\eea
Based on (\ref{Kabsdef}) the conformal weight $h_\cX$ of a holomorphic operator $\cX$ in the CFT is equivalent to $|K|$ in the gapless harmonic chain.
We compare the short interval expansion of the Schatten distance in the 2D free massless bosonic theory and that in the gapless limit of the harmonic chain in the left panels of figure~\ref{SchattenGaplessGapped}.
In the figure we only show the results for the holomorphic operators.
We find good matches when the interval is short.

When the conformal weights of the two states are the same, i.e. that $\d h =\d\bar h=0$, the distances $D_2( \r_{A,\cX_1}, \r_{A,\cX_2} )$ and $D_4( \r_{A,\cX_1}, \r_{A,\cX_2} )$ (\ref{D2D4}) would be vanishing at order $x^2$ and the leading order in the short interval expansion would be at least of order $x^4$.
This could be checked easily numerically in the harmonic chain, but we will not show it here.
This could also be calculated analytically in the massless bosonic theory, but we will not consider it in the present paper.

In the extremely massive limit of the bosonic theory, i.e. the extremely gapped limit of the harmonic chain, we get the universal short interval expansion of the Schatten distance \cite{ZRDistance}
\bea \label{D2D4X}
&& D_2( \r_{A,G}, \r_{A,\cX} ) = R x + o(x), \nn\\
&& D_4( \r_{A,G}, \r_{A,\cX} ) = R x + o(x),
\eea
with the total number of the excited quasiparticles $R$ defined as (\ref{Rdef}).
We compare the analytical short interval result with the numerical ones in the right panels of figure~\ref{SchattenGaplessGapped}.
There are good matches in the short interval regime.
Especially, when $R=1$, i.e. that only one quasiparticle is excited $K=k$, there are the exact results
\be
D_2( \r_{A,G}, \r_{A,k} ) = D_4( \r_{A,G}, \r_{A,k} ) = x.
\ee
Some preliminary results have been reported in \cite{Zhang:2020ouz} and we will report more details in the upcoming work \cite{ZRDistance}.

In figure~\ref{RenyiGaplessGapped} and figure~\ref{SchattenGaplessGapped}, it is interesting to note that in the gapless limit the short interval R\'enyi entropy and Schatten distance for the holomorphic states depend on $|K|$ defined in (\ref{Kabsdef}) and in the extremely massive limit they depend on $R$ defined in (\ref{Rdef}).

\section{Discussion}\label{secDis}

In this paper compared with the previous work \cite{Zhang:2020ouz}, we have considered the R\'enyi entropy and the Schatten distance in the 2D free massless non-compact bosonic field theory and the gapless limit of the harmonic chain in more complicated states.
Generally, the excited states in the harmonic chain have more than one quasiparticles, and thus we have called them multi-particle states.
We have found perfect matches of the analytical results in field theory with the numerical ones in the harmonic chain.
In the 2D free massless non-compact bosonic field theory we have only considered the states generated by the operators that could be constructed by the current operator, however, there are also the states generated by the vertex operators and their descendants, which are related to the zero mode in the gapless harmonic chain.
We hope to come back to the investigation of the R\'enyi entropy and subsystem Schatten distance involving the vertex operator states later \cite{ZRVertexOperator}.

In 2D free massless non-compact bosonic field theory we have calculated the universal leading order short interval expansion of the R\'enyi entropy (\ref{RenyiLeadingUniversal}) and Schatten distance (\ref{D2D4}), and in the extremely massive theory we found the universal leading order results (\ref{RenyiLeadingUniversal2}) and (\ref{D2D4X}).
In the gapless limit the leading order short interval R\'enyi entropy and Schatten distance depend on the energies of the states, while in the extremely gapped limit the leading order R\'enyi entropy and Schatten distance depend on the numbers of the excited quasiparticles. We think that similar behaviours should be also the case for more general interacting CFTs and QFTs.
It would be interesting to look for universal behaviors of the R\'enyi entropy and Schatten distance in different states for an interval with a finite length.

In \cite{Zhang:2020ouz} and the present paper, we have focused on the massless field theory and the gapless limit of the harmonic chain.
It is interesting to consider the R\'enyi entropy and the Schatten distance in the gapped, especially the extremely gapped, models.
The excited state R\'enyi entropy in the extremely gapped models under certain limit takes a universal form \cite{Castro-Alvaredo:2018dja,Castro-Alvaredo:2018bij,Castro-Alvaredo:2019irt,Castro-Alvaredo:2019lmj}, independent of the model, the quasi-particle momenta, and the connectedness of the subsystem.
 There exist additional contributions to the universal R\'enyi entropy when the momentum differences of quasiparticles are small.
We have reported the details for the R\'enyi entropy in \cite{Zhang:2020vtc,Zhang:2020dtd}.
Similar effect exists also for the Schatten distance.
We will report the universal subsystem Schatten distance and its corrections in \cite{ZRDistance}.

In the 2D free massless bosonic theory we have calculated analytically the second, the third and the fourth R\'enyi entropies and the second and the fourth Schatten distances.
It would be interesting to calculate the $n$-th R\'enyi entropy with general integer $n$ and the $n$-th Schatten distance with general even integer $n$ so that one could take the $n\to1$ limit to get the entanglement entropy and the trace distance.
This has been done for the entanglement entropies in the excited states of the vertex operator \cite{Alcaraz:2011tn,Berganza:2011mh} and the current operator \cite{Essler2013ShellFilling,Calabrese:2014Entanglement}, as well as for the trace distance among some special vertex operators \cite{Zhang:2019wqo,Zhang:2019itb}.
However, this would be difficult for the other excited states we have considered in this paper.
In most of the cases, we do not have the explicit forms of the general multi-point correlations functions, and even with the multi-point correlation functions known, it is still difficult to obtain the closed form of the general $n$-th R\'enyi entropy and the general $n$-th Schatten distance.
In the harmonic chain, it is also difficult to calculate numerically the excited state entanglement entropy and the trace distance, because the Hilbert space of a subsystem is always of an infinite dimension and it is difficult to find a basis to diagonalize the excited state RDMs.
We hope the excited state entanglement entropy and the trace distance in the free bosonic field theory and the harmonic chain can be calculated in the future.

\section*{Acknowledgements}

We thank Pasquale Calabrese and Sara Murciano for helpful discussions.
MAR thanks CNPq and FAPERJ (grant number 210.354/2018) for partial support.
JZ acknowledges support from ERC under Consolidator grant number 771536 (NEMO).

\appendix

\section{Derivation of the four-point function $\lag\cK\cK\cK\cK\rag_\rC$}\label{appKKKK}

In this appendix, we show two extra methods to calculate the four-point correlation function $F_{\cK\cK\cK\cK} (r)$ in (\ref{FXYZWr}).

In the first method, we do not need to know any information about the holomorphic primary operators with conformal weights larger than 3.
We first use the $TJ$ and the $T\cK$ OPEs in (\ref{OPEs}) and the four point correlation functions $\lag J(z_1)J(z_2)\cK(z_3)\cK(z_4)\rag$ and $\lag J(z_1)\cK(z_2)\cK(z_3)\cK(z_4)\rag$ in (\ref{FXYZWr}) and get the five-point correlation function $\lag T(z_1)J(z_2)\cK(z_3)\cK(z_4)\cK(z_5)\rag$, from which we get $\lag(TJ)(z_1)\cK(z_2)\cK(z_3)\cK(z_4)\rag$. Noting that $\cK=(TJ)-\f12\p^2J$, we combine $\lag(TJ)(z_1)\cK(z_2)\cK(z_3)\cK(z_4)\rag$ and $\lag\p^2J(z_1)\cK(z_2)\cK(z_3)\cK(z_4)\rag$ and obtain the four-point correlation function $\lag\cK(z_1)\cK(z_2)\cK(z_3)\cK(z_4)\rag$ consistent with $F_{\cK\cK\cK\cK} (r)$ in (\ref{FXYZWr}).

The second method is based on using the result of holomorphic bootstrap \cite{Headrick:2015gba}
\be \label{FKKKKrHB}
F_{\cK\cK\cK\cK} (r) = \f{c_0 (1 - r + r^2)^6 + c_1 r^2 (1 - r)^2 (1 - r + r^2)^3 + c_2 r^4(1 - r)^4}{(1 - r)^6}.
\ee
The coefficients $c_0,c_1,c_2$ can be determined by the expansion of the four-point correlation functions in terms of the global conformal block
\be \label{FKKKrGCB}
F_{\cK\cK\cK\cK} (r) = \sum_\cX \f{C_{\cK\cK\cX}^2}{\a_\cX} r^{h_\cX} {}_2F_1(h_\cX,h_\cX;2h_\cX;r),
\ee
where the sum $\cX$ is over all the holomorphic quasi-primary operators and ${}_2F_1$ is the hypergeometric function.
In small $r$ expansion of (\ref{FKKKrGCB}) to order $O(r^4)$, we need to sum $\cX=G,T,\cA,\cB$ with $h_\cX=0,2,4,4$, and then we compare it with (\ref{FKKKKrHB}) and get
\be
c_0 = \f94, ~~ c_1=c_2=27.
\ee
This reproduces $F_{\cK\cK\cK\cK} (r)$ in (\ref{FXYZWr}).

\section{Field theory results of the RDM products}\label{appPro}

In this appendix, we collect the results of the RDM products in the 2D free massless bosonic theory that are omitted in section~\ref{secCFT}.
We obtain the products of two RDMs
\be
\cF_\cA^{J,J} = \f{1}{128} [99 + 28 \cos(2 \pi x) + \cos(4 \pi x)],
\ee
\be
\cF_\cA^{\p J,\p J} = \f{1}{32768} [22931 + 8072 \cos(2 \pi x) + 1628 \cos(4 \pi x) +
    56 \cos(6 \pi x) + 81 \cos(8 \pi x)],
\ee
\be
\cF_\cA^{T,T} = \f{1}{131072} [88123 + 37256 \cos(2 \pi x) + 4604 \cos(4 \pi x) +
    1080 \cos(6 \pi x) + 9 \cos(8 \pi x)],
\ee
\bea
&& \cF_\cA^{\p^2 J,\p^2 J} = \f{1}{524288} [358254 + 119016 \cos(2 \pi x) + 34431 \cos(4 \pi x) +
    11044 \cos(6 \pi x) \nn\\
&& \phantom{\cF_\cA^{\p^2 J,\p^2 J} =}
    + 930 \cos(8 \pi x)- 12 \cos(10 \pi x) +
    625 \cos(12 \pi x)],
\eea
\bea
&& \cF_\cA^{\p T,\p T} = \f{1}{8388608} [5138606 + 2640936 \cos(2 \pi x) + 457599 \cos(4 \pi x) +
    114500 \cos(6 \pi x) \nn\\
&& \phantom{\cF_\cA^{\p T,\p T} =}
    + 15714 \cos(8 \pi x) + 20628 \cos(10 \pi x) +
    625 \cos(12 \pi x)],
\eea
\bea
&& \cF_\cA^{\cK,\cK} = \f{1}{33554432} [20470878 + 10480872 \cos(2 \pi x) + 1928535 \cos(4 \pi x) +
 554020 \cos(6 \pi x) \nn\\
&& \phantom{\cF_\cA^{\cK,\cK} =}
 + 113202 \cos(8 \pi x) + 6900 \cos(10 \pi x) + 25 \cos(12 \pi x)],
\eea
\be
\cF_\cA^{G,J} = \f{1}{2} [1 + \cos(\pi x)],
\ee
\be
\cF_\cA^{G,\p J} = \f{1}{16} [8 + 7 \cos(\pi x) + \cos(3 \pi x)],
\ee
\be
\cF_\cA^{G,T} = \f{1}{256} [99 + 128 \cos(\pi x) + 28 \cos(2 \pi x) + \cos(4 \pi x)],
\ee
\be
\cF_\cA^{G,\p^2 J} = \f{1}{128} [64 + 54 \cos(\pi x) + 7 \cos(3 \pi x) + 3 \cos(5 \pi x)],
\ee
\be
\cF_\cA^{G,\p T} = \f{1}{256} [106 + 120 \cos(\pi x) + 15 \cos(2 \pi x) + 8 \cos(3 \pi x) +
    6 \cos(4 \pi x) + \cos(6 \pi x)],
\ee
\be
\cF_\cA^{G,\cK} = \f{1}{1024} [338 + 486 \cos(\pi x) + 168 \cos(2 \pi x) + 23 \cos(3 \pi x) +
    6 \cos(4 \pi x) + 3 \cos(5 \pi x)],
\ee
\be
\cF_\cA^{J,\p J} = \f{1}{256} [106 + 120 \cos(\pi x) + 15 \cos(2 \pi x) + 8 \cos(3 \pi x) +
    6 \cos(4 \pi x) + \cos(6 \pi x)],
\ee
\be
\cF_\cA^{J,T} = \f{1}{1024} [454 + 434 \cos(\pi x) + 56 \cos(2 \pi x) + 69 \cos(3 \pi x) +
    2 \cos(4 \pi x) + 9 \cos(5 \pi x)],
\ee
\bea
&& \cF_\cA^{J,\p^2 J} = \f{1}{32768} [12161 + 15104 \cos(\pi x) + 3640 \cos(2 \pi x) +
    896 \cos(3 \pi x) \nn\\
&& \phantom{\cF_\cA^{J,\p^2 J} =}
    + 244 \cos(4 \pi x) + 384 \cos(5 \pi x) +
    264 \cos(6 \pi x) + 75 \cos(8 \pi x)],
\eea
\bea
&& \cF_\cA^{J,\p T} = \f{1}{4096} [1584 + 1747 \cos(\pi x) + 448 \cos(2 \pi x) + 233 \cos(3 \pi x) \nn\\
&& \phantom{\cF_\cA^{J,\p T} =}
  + 16 \cos(4 \pi x) + 43 \cos(5 \pi x) + 25 \cos(7 \pi x)],
\eea
\bea
&& \cF_\cA^{J,\cK} = \f{1}{65536} [23401 + 29440 \cos(\pi x) + 7640 \cos(2 \pi x) +
    2944 \cos(3 \pi x) \nn\\
&& \phantom{\cF_\cA^{J,\cK} =}
    + 1364 \cos(4 \pi x) + 384 \cos(5 \pi x) +
    360 \cos(6 \pi x) + 3 \cos(8 \pi x)],
\eea
\bea
&& \cF_\cA^{\p J,T} = \f{1}{8192} [2928 + 3795 \cos(\pi x) + 928 \cos(2 \pi x) + 233 \cos(3 \pi x) \nn\\
&& \phantom{\cF_\cA^{\p J,T} =}
    + 208 \cos(4 \pi x) + 43 \cos(5 \pi x) + 32 \cos(6 \pi x) +
    25 \cos(7 \pi x)],
\eea
\bea
&& \cF_\cA^{\p J,\p^2 J} = \f{1}{16384} [6482 + 7040 \cos(\pi x) + 1078 \cos(2 \pi x) +
    960 \cos(3 \pi x) + 424 \cos(4 \pi x) \nn\\
&& \phantom{\cF_\cA^{\p J,\p^2 J} =}
    + 192 \cos(5 \pi x) +
    175 \cos(6 \pi x) + 6 \cos(8 \pi x) + 27 \cos(10 \pi x)],
\eea
\bea
&& \hspace{-5mm} \cF_\cA^{\p J,\p T} = \f{1}{131072} [45862 + 57358 \cos(\pi x) + 16144 \cos(2 \pi x) +
    5412 \cos(3 \pi x) + 3256 \cos(4 \pi x) \nn\\
&& \hspace{-5mm} \phantom{\cF_\cA^{\p J,\p T} =}
    + 2260 \cos(5 \pi x) +
    112 \cos(6 \pi x) + 217 \cos(7 \pi x) + 162 \cos(8 \pi x) +
    289 \cos(9 \pi x)],
\eea
\bea
&& \cF_\cA^{\p J,\cK} = \f{1}{131072} [41746 + 58696 \cos(\pi x) + 19718 \cos(2 \pi x) +
    5016 \cos(3 \pi x) + 3368 \cos(4 \pi x) \nn\\
&& \phantom{\cF_\cA^{\p J,\cK} =}
    + 1224 \cos(5 \pi x) +
    503 \cos(6 \pi x) + 600 \cos(7 \pi x) + 198 \cos(8 \pi x) +
    3 \cos(10 \pi x)],
\eea
\bea
&& \cF_\cA^{T,\p^2 J} = \f{1}{32768} [10305 + 14748 \cos(\pi x) + 5432 \cos(2 \pi x) +
    1200 \cos(3 \pi x) + 308 \cos(4 \pi x) \nn\\
&& \phantom{\cF_\cA^{T,\p^2 J} =}
    + 320 \cos(5 \pi x) +
    264 \cos(6 \pi x) + 62 \cos(7 \pi x) + 75 \cos(8 \pi x) +
    54 \cos(9 \pi x)],
\eea
\bea
&& \cF_\cA^{T,\p T} = \f{1}{131072} [46390 + 56088 \cos(\pi x) + 15122 \cos(2 \pi x) +
    8328 \cos(3 \pi x) + 2936 \cos(4 \pi x) \nn\\
&& \phantom{\cF_\cA^{T,\p T} =}
    + 920 \cos(5 \pi x) +
    485 \cos(6 \pi x) + 200 \cos(7 \pi x) + 594 \cos(8 \pi x) +
    9 \cos(10 \pi x)],
\eea
\bea
&& \cF_\cA^{T,\cK} = \f{1}{524288} [213714 + 198042 \cos(\pi x) + 43952 \cos(2 \pi x) +
    51276 \cos(3 \pi x) + 3752 \cos(4 \pi x) \nn\\
&& \phantom{\cF_\cA^{T,\cK} =}
    + 10076 \cos(5 \pi x)
    + 720 \cos(6 \pi x) + 2675 \cos(7 \pi x)
    + 6 \cos(8 \pi x)
    + 75 \cos(9 \pi x)],
\eea
\bea
&& \cF_\cA^{\p^2 J,\p T} = \f{1}{1048576} [356944 + 449538 \cos(\pi x) + 123456 \cos(2 \pi x) +
    59134 \cos(3 \pi x) \nn\\
&& \phantom{\cF_\cA^{\p^2 J,\p T} =}
    + 29760 \cos(4 \pi x) + 11463 \cos(5 \pi x) +
    11872 \cos(6 \pi x) + 2649 \cos(7 \pi x) \nn\\
&& \phantom{\cF_\cA^{\p^2 J,\p T} =} + 1392 \cos(8 \pi x) +
    181 \cos(9 \pi x) + 864 \cos(10 \pi x) + 1323 \cos(11 \pi x)],
\eea
\bea
&& \cF_\cA^{\p^2 J,\cK} = \f{1}{16777216} [4729230 + 7246848 \cos(\pi x) + 3247464 \cos(2 \pi x) +
    880640 \cos(3 \pi x) \nn\\
&& \phantom{\cF_\cA^{\p^2 J,\cK} =}
    + 229647 \cos(4 \pi x) + 172032 \cos(5 \pi x) +
    130148 \cos(6 \pi x) + 47616 \cos(7 \pi x) \nn\\
&& \phantom{\cF_\cA^{\p^2 J,\cK} =} + 39042 \cos(8 \pi x) +
    41472 \cos(9 \pi x) + 12852 \cos(10 \pi x) + 225 \cos(12 \pi x)],
\eea
\bea
&& \cF_\cA^{\p T,\cK} = \f{1}{2097152} [674000 + 890994 \cos(\pi x) + 286272 \cos(2 \pi x) +
    127694 \cos(3 \pi x) \nn\\
&& \phantom{\cF_\cA^{\p T,\cK} =}
    + 68160 \cos(4 \pi x) + 22191 \cos(5 \pi x) +
    13664 \cos(6 \pi x) + 2577 \cos(7 \pi x) \nn\\
&& \phantom{\cF_\cA^{\p T,\cK} =}+ 6384 \cos(8 \pi x) +
    4877 \cos(9 \pi x) + 96 \cos(10 \pi x) + 243 \cos(11 \pi x)],
\eea
the products of three RDMs
\be
\cF_\cA^{J,J,J} = \f{1}{81} [51 + 28 \cos(2\pi x) + 2 \cos(4\pi x) ],
\ee
\bea
&& \cF_\cA^{\p J,\p J,\p J} = \f{1}{177147} [94401 + 64212 \cos(2\pi x) + 15852 \cos(4\pi x) \nn\\
&& \phantom{\cF_\cA^{\p J,\p J,\p J} =}
  + 1908 \cos(6\pi x) + 742 \cos(8\pi x) + 32 \cos(10\pi x) ],
\eea
and the products of four RDMs
\bea
&& \cF_\cA^{J,J,J,J} = \f{1}{8388608} [4483347 + 3477832 \cos(2\pi x) + 416476 \cos(4\pi x) \nn\\
&& \phantom{\cF_\cA^{J,J,J,J} =}
  + 10872 \cos(6\pi x) + 81 \cos(8\pi x) ],
\eea
\bea
&& \cF_\cA^{\p J,\p J,\p J,\p J} = \f{1}{140737488355328} [60874090392995 + 57949920208208 \cos(2\pi x) \nn\\
&& \phantom{\cF_\cA^{\p J,\p J,\p J,\p J} =}
  + 17516408426312 \cos(4\pi x) + 3372546127088 \cos(6\pi x) \nn\\
&& \phantom{\cF_\cA^{\p J,\p J,\p J,\p J} =} +
   912235768092 \cos(8\pi x) + 99964047312 \cos(10\pi x) \nn\\
&& \phantom{\cF_\cA^{\p J,\p J,\p J,\p J} =} +
   11330531448 \cos(12\pi x) + 861591024 \cos(14\pi x) \nn\\
&& \phantom{\cF_\cA^{\p J,\p J,\p J,\p J} =} +
   131262849 \cos(16\pi x) ],
\eea
\be
\cF_\cA^{G,G,G,J} = \f{1}{8}\Big[2 + 3 \cos\f{\pi x}{2} + 2 \cos(\pi x) + \cos\f{3\pi x}{2}\Big],
\ee
\bea
&& \cF_\cA^{G,G,J,J} = \f{1}{512}\Big[99 + 160 \cos\f{\pi x}{2} + 120 \cos(\pi x) +
   80 \cos\f{3\pi x}{2} + 28 \cos(2\pi x) \nn\\
&& \phantom{\cF_\cA^{G,G,J,J} =}
   + 16 \cos\f{5\pi x}{2} +
   8 \cos(3\pi x) + \cos(4\pi x)\Big],
\eea
\be
\cF_\cA^{G,J,G,J} = \f{1}{2048} [707 + 992 \cos(\pi x) + 316 \cos(2\pi x) + 32 \cos(3\pi x) +
 \cos(4\pi x) ],
\ee
\bea
&& \cF_\cA^{G,J,J,J} = \f{1}{32768}\Big[6124 + 8142 \cos\f{\pi x}{2} + 6948 \cos(\pi x) +
 5882 \cos\f{3\pi x}{2} + 1904 \cos(2\pi x) \nn\\
&& \phantom{\cF_\cA^{G,J,J,J} =}
 + 2037 \cos\f{5\pi x}{2} + 1226 \cos(3\pi x) +
 223 \cos\f{7\pi x}{2} + 164 \cos(4\pi x) \nn\\
&& \phantom{\cF_\cA^{G,J,J,J} =}
 + 99 \cos\f{9\pi x}{2} +
 18 \cos(5\pi x) + \cos\f{11\pi x}{2} \Big],
\eea
\be
\cF_\cA^{G,G,G,\p J} = \f{1}{256}\Big[64 + 99 \cos\f{\pi x}{2} + 56 \cos(\pi x) +
   17 \cos\f{3\pi x}{2} + 3 \cos\f{5\pi x}{2} + 8 \cos(3\pi x) +
   9 \cos\f{7\pi x}{2}\Big],
\ee
\bea
&& \cF_\cA^{G,G,\p J,\p J} = \f{1}{131072}\Big[22931 + 38528 \cos\f{\pi x}{2} + 26646 \cos(\pi x) +
   16048 \cos\f{3\pi x}{2} \nn\\
&& \phantom{\cF_\cA^{G,G,\p J,\p J} =}
   + 8072 \cos(2\pi x) +
   5264 \cos\f{5\pi x}{2} + 5330 \cos(3\pi x) +
   4224 \cos\f{7\pi x}{2} \nn\\
&& \phantom{\cF_\cA^{G,G,\p J,\p J} =}
   + 1628 \cos(4\pi x) +
   1280 \cos\f{9\pi x}{2} + 630 \cos(5\pi x) +
   48 \cos\f{11\pi x}{2} \nn\\
&& \phantom{\cF_\cA^{G,G,\p J,\p J} =}
   + 56 \cos(6\pi x) +
   144 \cos\f{13\pi x}{2} + 162 \cos(7\pi x) + 81 \cos(8\pi x)\Big],
\eea
\bea
&& \cF_\cA^{G,\p J,G,\p J} = \f{1}{8388608} [2764003 + 3540352 \cos(\pi x) + 1062344 \cos(2\pi x)\nn\\
&& \phantom{\cF_\cA^{G,\p J,G,\p J} =}
 + 603264 \cos(3\pi x)
 + 352028 \cos(4\pi x) + 40320 \cos(5\pi x) \nn\\
&& \phantom{\cF_\cA^{G,\p J,G,\p J} =}
 + 14840 \cos(6\pi x) + 10368 \cos(7\pi x) +
 1089 \cos(8\pi x) ],
\eea
\bea
&& \cF_\cA^{G,\p J,\p J,\p J} = \f{1}{4294967296}\Big[727742848 + 851247918 \cos\f{\pi x}{2} +
   760154160 \cos(\pi x) \nn\\
&& \phantom{\cF_\cA^{G,\p J,\p J,\p J} =}
   + 733713626 \cos\f{3\pi x}{2} +
   265235456 \cos(2\pi x) + 304151358 \cos\f{5\pi x}{2} \nn\\
&& \phantom{\cF_\cA^{G,\p J,\p J,\p J} =} +
   250361296 \cos(3\pi x) + 139896826 \cos\f{7\pi x}{2} +
   71347712 \cos(4\pi x) \nn\\
&& \phantom{\cF_\cA^{G,\p J,\p J,\p J} =} + 83521773 \cos\f{9\pi x}{2} +
   54728616 \cos(5\pi x) + 13981711 \cos\f{11\pi x}{2} \nn\\
&& \phantom{\cF_\cA^{G,\p J,\p J,\p J} =} +
   4419584 \cos(6\pi x) + 16433421 \cos\f{13\pi x}{2} +
   7563864 \cos(7\pi x) \nn\\
&& \phantom{\cF_\cA^{G,\p J,\p J,\p J} =} + 2394775 \cos\f{15\pi x}{2} +
   4913280 \cos(8\pi x) + 1728267 \cos\f{17\pi x}{2} \nn\\
&& \phantom{\cF_\cA^{G,\p J,\p J,\p J} =} +
   835320 \cos(9\pi x) + 120969 \cos\f{19\pi x}{2} +
   82944 \cos(10\pi x) \nn\\
&& \phantom{\cF_\cA^{G,\p J,\p J,\p J} =} + 283203 \cos\f{21\pi x}{2} +
   98568 \cos(11\pi x) + 9801 \cos\f{23\pi x}{2}\Big],
\eea
\bea
&& \cF_\cA^{J,J,J,\p J} = \f{1}{16777216}\Big[2459038 + 3883780 \cos\f{\pi x}{2} + 3247180 \cos(\pi x) +
   2499128 \cos\f{3\pi x}{2} \nn\\
&& \phantom{\cF_\cA^{J,J,J,\p J} =}
   + 1408306 \cos(2\pi x) +
   1239016 \cos\f{5\pi x}{2} + 866936 \cos(3\pi x) +
   544384 \cos\f{7\pi x}{2} \nn\\
&& \phantom{\cF_\cA^{J,J,J,\p J} =}
   + 223480 \cos(4\pi x) +
   168656 \cos\f{9\pi x}{2} + 78008 \cos(5\pi x) +
   40504 \cos\f{11\pi x}{2} \nn\\
&& \phantom{\cF_\cA^{J,J,J,\p J} =}
   + 97405 \cos(6\pi x) +
   12456 \cos\f{13\pi x}{2} + 2162 \cos(7\pi x) +
   522 \cos\f{15\pi x}{2} \nn\\
&& \phantom{\cF_\cA^{J,J,J,\p J} =}
   + 5994 \cos(8\pi x) +
   162 \cos\f{17\pi x}{2} + 18 \cos(9\pi x) + 81 \cos(10\pi x)\Big],
\eea
\bea
&& \cF_\cA^{J,J,\p J,\p J} = \f{1}{536870912}\Big[71314222 + 118993344 \cos\f{\pi x}{2} +
   99151152 \cos(\pi x) + 80648096 \cos\f{3\pi x}{2} \nn\\
&& \phantom{\cF_\cA^{J,J,\p J,\p J} =} +
   47802792 \cos(2\pi x) + 39463968 \cos\f{5\pi x}{2} +
   30125904 \cos(3\pi x) + 16831872 \cos\f{7\pi x}{2} \nn\\
&& \phantom{\cF_\cA^{J,J,\p J,\p J} =} +
   12232191 \cos(4\pi x) + 7373184 \cos\f{9\pi x}{2} +
   4561128 \cos(5\pi x) + 3059664 \cos\f{11\pi x}{2} \nn\\
&& \phantom{\cF_\cA^{J,J,\p J,\p J} =} +
   2478340 \cos(6\pi x) + 1320336 \cos\f{13\pi x}{2} +
   366936 \cos(7\pi x) + 580256 \cos\f{15\pi x}{2} \nn\\
&& \phantom{\cF_\cA^{J,J,\p J,\p J} =} +
   330978 \cos(8\pi x) + 146592 \cos\f{17\pi x}{2} +
   12536 \cos(9\pi x) + 14544 \cos\f{19\pi x}{2} \nn\\
&& \phantom{\cF_\cA^{J,J,\p J,\p J} =} +
   57684 \cos(10\pi x) + 3600 \cos\f{21\pi x}{2} +
   72 \cos(11\pi x) + 1521 \cos(12\pi x)\Big],
\eea
\bea
&& \cF_\cA^{J,\p J,J,\p J} = \f{1}{34359738368} [8265067022 + 13131207360 \cos(\pi x) +
   7069516392 \cos(2\pi x) \nn\\
&& \phantom{\cF_\cA^{J,\p J,J,\p J} =}
   + 3313666880 \cos(3\pi x) +
   1556602575 \cos(4\pi x) + 611651232 \cos(5\pi x) \nn\\
&& \phantom{\cF_\cA^{J,\p J,J,\p J} =} +
   262197476 \cos(6\pi x) + 119162976 \cos(7\pi x) +
   23615490 \cos(8\pi x) \nn\\
&& \phantom{\cF_\cA^{J,\p J,J,\p J} =}
   + 4145888 \cos(9\pi x) +
   2795700 \cos(10\pi x) + 34848 \cos(11\pi x) \nn\\
&& \phantom{\cF_\cA^{J,\p J,J,\p J} =} +
   74529 \cos(12\pi x) ],
\eea
\bea
&& \cF_\cA^{J,\p J,\p J,\p J} = \f{1}{68719476736}\Big[9258218916 + 13744769768 \cos\f{\pi x}{2} +
   12255718376 \cos(\pi x) \nn\\
&& \phantom{\cF_\cA^{J,\p J,\p J,\p J} =}
   + 10474165728 \cos\f{3\pi x}{2} +
   5454263579 \cos(2\pi x) + 5617875584 \cos\f{5\pi x}{2} \nn\\
&& \phantom{\cF_\cA^{J,\p J,\p J,\p J} =} +
   4011577022 \cos(3\pi x) + 2414479790 \cos\f{7\pi x}{2} +
   1974982682 \cos(4\pi x) \nn\\
&& \phantom{\cF_\cA^{J,\p J,\p J,\p J} =}
   + 1206433510 \cos\f{9\pi x}{2} +
   775483078 \cos(5\pi x) + 555879200 \cos\f{11\pi x}{2} \nn\\
&& \phantom{\cF_\cA^{J,\p J,\p J,\p J} =} +
   365274777 \cos(6\pi x) + 212485200 \cos\f{13\pi x}{2} +
   116487676 \cos(7\pi x) \nn\\
&& \phantom{\cF_\cA^{J,\p J,\p J,\p J} =}
   + 93041756 \cos\f{15\pi x}{2} +
   116775036 \cos(8\pi x) + 30024060 \cos\f{17\pi x}{2} \nn\\
&& \phantom{\cF_\cA^{J,\p J,\p J,\p J} =}
 + 19652956 \cos(9\pi x) + 8514384 \cos\f{19\pi x}{2} +
   8516043 \cos(10\pi x) \nn\\
&& \phantom{\cF_\cA^{J,\p J,\p J,\p J} =}
 + 1721568 \cos\f{21\pi x}{2} +
   946026 \cos(11\pi x) + 285162 \cos\f{23\pi x}{2} \nn\\
&& \phantom{\cF_\cA^{J,\p J,\p J,\p J} =} +
   1735110 \cos(12\pi x) + 62658 \cos\f{25\pi x}{2} +
   4050 \cos(13\pi x) \nn\\
&& \phantom{\cF_\cA^{J,\p J,\p J,\p J} =}
 + 103041 \cos(14\pi x)\Big].
\eea

\providecommand{\href}[2]{#2}\begingroup\raggedright\endgroup


\end{document}